\documentclass[aps, amsmath,amssymb,pra,twocolumn,superscriptaddress]{revtex4-2}

\usepackage{graphicx}
\usepackage{dcolumn}
\usepackage[dvipsnames]{xcolor}
\usepackage{bm}

\usepackage{empheq,etoolbox}
\newcommand{\revisionB}[1]{\textcolor{magenta}{#1}}

\begin{document}


\title{
Stable vortex structures in colliding self-gravitating Bose-Einstein condensates}
\author{Y.O. Nikolaieva}
\affiliation{Department of Physics, Taras Shevchenko National University of Kyiv, 
64/13, Volodymyrska Street, Kyiv 01601, Ukraine}
\affiliation{Physikalisch-Technische Bundesanstalt (PTB), Bundesallee 100, D-38116 Braunschweig, Germany}
\author{Y.M. Bidasyuk}
\affiliation{Physikalisch-Technische Bundesanstalt (PTB), Bundesallee 100, D-38116 Braunschweig, Germany}
\author{K. Korshynska}
\affiliation{Department of Physics, Taras Shevchenko National University of Kyiv, 
64/13, Volodymyrska Street, Kyiv 01601, Ukraine}
\affiliation{Physikalisch-Technische Bundesanstalt (PTB), Bundesallee 100, D-38116 Braunschweig, Germany}
\author{E.V. Gorbar}
\affiliation{Department of Physics, Taras Shevchenko National University of Kyiv, 
64/13, Volodymyrska Street, Kyiv 01601, Ukraine}
\affiliation{Bogolyubov Institute for Theoretical Physics, 14-b Metrolohichna Street, Kyiv 03143, Ukraine}
\author{Junji Jia}
\affiliation{School of Physics and Technology, Wuhan University, 
299 Bayi Road, Wuhan, 
Hubei Prov., China 430072}
\author{A.I. Yakimenko}
\affiliation{Department of Physics, Taras Shevchenko National University of Kyiv,
64/13, Volodymyrska Street, Kyiv 01601, Ukraine}
\affiliation{Dipartimento di Fisica e Astronomia ’Galileo Galilei’,
Universit{\'a} di Padova, via Marzolo 8, 35131 Padova, Italy}

\begin{abstract} 
A key feature of ultra-light dark matter composed by bosons is {the formation of} superfluid Bose-Einstein condensate (BEC) structures on galactic scales. 
We study collisions of BEC solitonic and vortex structures in the framework of the Gross-Pitaevskii-Poisson model. It is found that the superfluid nature of bosonic dark matter leads to the formation of quantized vortex lines and vortex rings in interference patterns formed during collisions.   Calculating the gravitational wave luminosity, we demonstrated that quantum interference patterns affect notably the gravitational wave radiation.
 We reveal that superfluid self-gravitating BECs can form stable localized vortex structures which remain robust even after  a head-on collision. 
\end{abstract}



\maketitle

\section{Introduction}
The nature, composition, and physical properties of dark matter (DM) are key puzzles of modern physics, astrophysics, and cosmology. DM composed by ultralight bosons in the state of Bose-Einstein condensate (BEC) presents an appealing possibility because it naturally resolves some problems of the standard $\Lambda$CDM model on the galactic scale while maintaining the success of the latter on larger scales (for a review, see \cite{2017NJPh...19b3019E}). These conclusions are well-supported by cosmological simulations \cite{2014NatPh..10..496S}. The large-scale structure of the ultra-light DM simulations is indistinguishable from CDM, being consistent with astrophysical observations. In cosmological studies, it was found that the boson mass has to be ultralight to reproduce the observed distribution of matter at large scales \cite{2001PhRvD..63f3506M, 2000PhRvD..62j3517S}. For such extremely small mass, namely, of the order of $m \sim 10^{-22}$ eV \cite{2014NatPh..10..496S}, quantum mechanical phenomena manifest themselves on galactic scales. The viability of the fuzzy DM model, a kind of ultra-light DM without self-interaction, was studied with the stellar kinematics measurements in dwarf galaxies \cite{Goldstein_2022}.

The evolution of a self-gravitating galactic BEC is governed by the system of the Gross-Pitaevskii-Poisson (GPP) equations. These equations appear in different areas of physics and describe such seemingly distinct physical processes as the nonlinear propagation of optical beams in nonlocal media, atomic BECs, and the evolution of cold DM galactic halos \cite{2020PhyD..40332301P, PhysRevA.81.063617}. This opens an intriguing possibility of modeling cosmological structures via their laboratory analogs, with the results obtained in one area can be applied in the others \cite{PhysRevLett.84.5687, PhysRevA.63.031603, 2001EL.....56....1G}.  BEC structures could play an important role in astrophysics and cosmology. They have been predicted to form the central cores of ultra-light axion DM halos. Similar proposals to describe DM structures as BEC were suggested in many studies \cite{2007JCAP...06..025B, 2017EPJP..132..248C, Chavanis_2012, Hui_2017}.

A well-established fact is that compact spherical astrophysical objects that may be formed due to the Bose-Einstein condensation of DM are stable, as was found numerically \cite{2015PhRvD..91d4041M, Siemonsen_2021}. Since BEC is a superfluid, it could form vortices with quantized angular momentum \cite{yakimenko2005stable, lashkin2009stable}. Rotation in BEC halos would affect their structure and thus could lead to observable consequences \cite{Rindler_Daller_2012, Sanchis_Gual_2019, zhang2018slowly}. Stationary vortex states of a spinning DM cloud with different topological charges and typical galactic halo mass and radius were recently studied in \cite{2021LTP....47..684N}. It was found that while all multi-charged vortex states are unstable, a single-charged vortex state is very robust, even being strongly perturbed, and survives during the lifetime of the Universe. An analysis of the stability of solutions with different topological charges and interaction constant values, e.g, bosonic stars, was performed in \cite{PhysRevD.104.023504} where it was found that only state with topological charge $s=0$ and $s=1$ could be stable. Furthermore, fundamental solitons form an interference pattern during a head-on collision, but survive and revive their forms after collision \cite{PhysRevA.66.063609, PhysRevD.94.063503, 2011PhRvD..83j3513G, 2006PhRvD..74j3002B, 2002PhRvA..66f3609C}. This raises a more sophisticated  question of whether the vortex solitons also remain stable after a head-on collision. To the best of our knowledge, stability of colliding vortex solitons in self-gravitating BEC has never been investigated.

An important clue to revealing DM nature is the first indication of DM nongravitational self-interaction, which has been recently reported for clusters collisions \cite{2010ApJ...715L.160C, Harvey_2015}. These observations could be explained by the ultra-light DM model with collisional dynamics of stable solitary solutions of the Schrödinger–Poisson equation \cite{2016PDU....12...50P}.  The nature of BECs severely affects the collisional dynamics of DM clumps and could explain some recent puzzling observations. 

Binary solitonic collisions were studied both analytically and numerically \cite{PhysRevD.94.063503}. These collisions are characterized by the relative amounts of the kinetic, self-interaction, and gravitational binding energies. Systems with “negative” energy (when the kinetic and self-interaction energies are smaller than the gravitational binding energy) will combine and merge, whereas systems with “positive” energy will behave as solitons and pass right through each other. The 3D cosmological simulations with merging multiple solitons to create individual virialized objects were conducted \cite{Harvey_2015, 2018PhRvD..97k6003G, 2016PhRvD..94d3513S, liu2022coherent}. During the merger of two-state configurations, the total density approaches a stationary state. The obtained averaged profile has a core plus a tail structure that could serve to explain the results in simulations of DM structure formation \cite{Harvey_2015, 2016PhRvD..94d3513S}. Similar numerical solutions were obtained in the case of collision of two spherical BEC cores, with \cite{2014PhRvL.113z1302S} and without \cite{Maleki_2020, 2016PhRvD..94d3513S} the nonlinear self-interaction term in the GPP equations. In collisions of solitonic cores with opposite phases, a destructive interference occurs, which gives rise to a short-range repulsive force between the cores \cite{Paredes_2016}. As the authors of \cite{Carrasco_2010} pointed out in the context of galaxy cluster observations with indications of an offset between dark and stellar matter \cite{Carrasco_2010, Massey_2015}, this effect can provide an alternative explanation to self-interacting DM.

The relaxation process during BEC structures collision suggests a rich phenomenology \cite{2019PhRvD..99d3542A}. Mergers of two solitons undergo relaxation in the form of gravitational cooling \cite{2016PhRvD..94d3513S}. The gravitational cooling process for initially quite arbitrary density profiles leads to relaxation and virialization through the emission of the bosonic particles \cite{PhysRevD.93.103535}. In addition, density oscillations take place and the period of these oscillations could range from a fraction of gigayear up to many gigayears \cite{2016PhRvD..94d3513S}.

Alternative treatment of the soliton-soliton collision and orbiting binary boson stars is discussed in \cite{2007PhRvD..75f4005P, 2008PhRvD..77d4036P, 2017PhRvD..95l4005B}, where instead of the standard non-relativistic GPP system of equations, the Einstein-Klein-Gordon system of equations is solved. The first studies of head-on collisions of mini bosonic stars were performed in \cite{1999PhDT........44B} by using 3D code. Ultrarelativistic collisions were considered in \cite{2010PhRvL.104k1101C}, and head-on and orbital mergers of non-identical boson stars in \cite{2007PhRvD..75f4005P, 2008PhRvD..77d4036P}.

Obviously, since only the luminous matter is directly observable, collisions of galaxies composed of dark and luminous matter could provide very important information about DM \cite{PhysRevD.93.103535}. Studies in this direction \cite{2016PhRvD..93j3535G} show that luminous matter cannot follow these extreme dynamics and is expelled from the gravitational potential. The Bullet Cluster gives the famous example \cite{2004ApJ...606..819M}, in which the visible matter and DM are spatially separated after the collision. Therefore, it is interesting to determine the characteristics of collisions of self-gravitating BECs in states with nonzero angular momentum too. This provides the motivation for the study in this paper.

One of the potentially observable characteristics of the collision process is the emission of gravitational waves (GW). It is worth mentioning that gravitational waves can allow us to find unexpected astrophysical compact objects with low brightness, known generically as Exotic Compact Objects (ECOs). Among the most plausible ECO candidates are the bosonic stars, self-gravitating objects made of a complex scalar field \cite{2012LRR....15....6L, 2013PhRvD..88f4046M}. Boson stars provide a simple and useful model to study compact bodies in different scenarios, such as DM candidates and black hole mimickers.

In this paper, we consider collisions of DM structures in the Bose-Einstein condensate state for different orientations of the angular momentum. We found that the BEC clouds after the formation of the interference pattern during the collision restore their topological structure even for an arbitrary impact parameter and the initial orientation of their vortex cores.

The paper is organized as follows. The model is described in Sec.\ref{sec:Model}. BEC states with nonzero angular momentum are considered in Sec.\ref{sec:stationary}. Binary collisions are studied in Sec.\ref{sec:binary}. The emission of gravitational waves  produced by colliding superfluid DM structures is investigated in Sec.\ref{sec:GW}. Conclusions are drawn in Sec.\ref{sec:conclusions}.

\section{Model}\label{sec:Model}

At zero temperature, the dynamics of self-gravitating BEC of weakly interacting bosons in the mean-field  approximation is described by the system of the Gross-Pitaevskii-Poisson  (GPP) equations for the BEC and gravitational potential, which in the dimensionless units takes the following form:
\begin{eqnarray}
 i\frac{\partial\Psi}{\partial t} = \left(-\frac{1}{2}\nabla^2 +\Phi +\nu |\Psi|^2 \right) \Psi,
 \label{GP-equation}\\
 \nabla^2 \Phi=|\Psi|^2,
 \label{Poisson-equation}
\end{eqnarray}
where $\Psi(\mathbf{r},t)$ is a complex wave function of the condensate, $\mathbf{r}=(x,y,z)$ is the vector of  spatial coordinates, and $t$ is time. The Gross-Pitaevskii equation (\ref{GP-equation}) governing $\Psi$ is a nonlinear Schrödinger equation with non-linearity sourced by gravitational potential $\Phi(\mathbf{r},t)$ and nonlocal interparticle interaction. Coefficient $\nu$ takes the value $+1$ for repulsive interparticle interaction, $-1$ for attractive interparticle interaction, and $0$ if particles do not interact. In this work, we consider the case of repulsive interparticle interaction $\nu=1$.
The total energy associated with the GPP system of equations can be written as 
\begin{equation}
\label{Etot}
E=\Theta +U+W,
\end{equation}
where, in the dimensionless units, the kinetic energy is
\begin{equation}\label{eq:E_kinetic}
    \Theta=\frac{1}{2}\int |\nabla\Psi|^2d\mathbf{r},
\end{equation}
the  internal energy 
\begin{equation}
    U = \frac{1}{2}\int |\Psi|^4d\mathbf{r},
\end{equation}
and the gravitational potential energy of interaction is given by
 \begin{equation}
     W =\frac{1}{2}\int |\Psi|^2\Phi d\mathbf{r}.
 \end{equation}



To obtain dimensional quantities the following relations can be used
$\mathbf{r}_\mathrm{ph}=\mathbf{r}L_*$,
$t_\mathrm{ph}= t/\Omega_* $, 
$\Phi_\mathrm{ph}= \Phi \phi_*$,
$\Psi_\mathrm{ph}= \Psi \psi_*$,
$E_\mathrm{ph}=E \epsilon_*$,
where $L_*=\lambda_C (m_{\textrm{Pl}}/m)\sqrt{\lambda/8\pi}$,
$\Omega_*=c\lambda_C/L_*^2$,
$\phi_*=(c\lambda_C/L_*)^2$,
$\psi_*=mc^2/\left((\lambda/8\pi)\left(m_\textrm{Pl}/m\right)^2\sqrt{4\pi G M}\hbar\right)$,
$\epsilon_* = \hbar^2\left(8\pi/\lambda\right)^{3/2}/(4\pi m_\textrm{Pl}\lambda_{C}^2)$, $m_{\textrm{Pl}}=\sqrt{\hbar c/G}$ is the Planck mass, $\lambda/8\pi=a_s/\lambda_{C}$ is the self-interaction constant with $a_s$ being the $s$-wave scattering length, $\lambda_{C}=\hbar/mc$ is the Compton wavelength of the bosons and $M$ is the total BEC cloud mass. 

The normalization constant $N_{\text{s}}$ in dimensionless units is as follows:
\begin{equation}
\label{eq:normalized_dimless}
  \int|\Psi|^2d\textbf{r}=\frac{1}{\psi_*^2L_*^3}=4\pi\frac{M}{m_{\textrm{Pl}}} \sqrt{\frac{\lambda}{8\pi}}=N_{\text{s}}.
\end{equation}
Further, we fix $N_s=100$, unless otherwise noted. 

Before proceeding to the study of BEC DM structures collision, let us consider the properties of stationary self-gravitating BEC solitons.

\section{Stationary states}
\label{sec:stationary}

As a result of the balance between gravitational attraction, quantum pressure, and repulsive bosonic interparticle interaction, the system of equations (\ref{GP-equation}) and (\ref{Poisson-equation}) allows the existence of stationary solutions with the wave function
\begin{equation}
\Psi(\mathbf{r},t) = \psi(\mathbf{r})e^{{-i\mu t}},   
\label{eq:stat_st}
\end{equation}
where $\mu$ is the chemical potential and $\psi(\mathbf{r})$ is the spatial profile of the wave function. 
We seek for  $\psi(\mathbf{r})$ in the form
\begin{equation*}
  \psi(r_{\perp},\theta,z) = \chi(r_{\perp},z)e^{{is\theta}},
\end{equation*}
where $r_{\perp} = \sqrt{x^2+y^2}$, and the integer number $s$ is the topological charge. 
In \cite{2021LTP....47..684N}, we have analytically and numerically analyzed the stationary states with different topological charges and investigated their stability. It has been found that the self-gravitating BEC states with $s=0$ (the fundamental soliton) and $s=1$ (the vortex soliton) can be stable even under the strong perturbations, what is also in good agreement with the results obtained in \cite{PhysRevD.104.023504}.
For this work, we have developed an efficient numerical scheme for solving the system of equations (\ref{GP-equation}), (\ref{Poisson-equation}) in three spatial dimensions. Stationary states were numerically obtained by evolving the system (\ref{GP-equation}), (\ref{Poisson-equation}) in imaginary time with a normalization constraint. We therefore can use the same numerical methods for stationary states and for the dynamics of the system. The Gross-Pitaevskii equation is solved using standard pseudo-spectral "split-step" method \cite{ANTOINE20132621}. For the Poisson equation, we adapt the well-known geometric multigrid method. While the Poisson equation does not explicitly depend on time, the values of gravitational potential at every time step are obtained using a single multigrid V-cycle with 19-point Jacobi smoother \cite{PhysRevE.52.1181,PhysRevD.105.083521,OReilly2006AFO}, and using the result from a previous time step as an initial condition for the next one.
Additionally, in order to keep the computational domain reasonably small, a careful consideration of boundary conditions for the Poisson equation is extremely important. We set the values of the potential $\Phi$ at the boundaries using a combination of monopole and quadrupole potentials 
$$
\Phi = -\frac{N_s}{4\pi r} - \frac{1}{8\pi r^5} \sum_{ij} I_{ij} x_i x_j, 
$$
where the quadrupole moment $I_{ij}$ 
\begin{equation}\label{eq:I_ij}
    I_{ij}=\int |\Psi|^2 \left(x_ix_j-\frac{1}{3}\delta_{i,j}x^2 \right) dx^3.
\end{equation}
is calculated from the current mass distribution and updated at every time step accordingly.

\begin{figure}[ht]
\begin{center}
\includegraphics[width=3.4in]{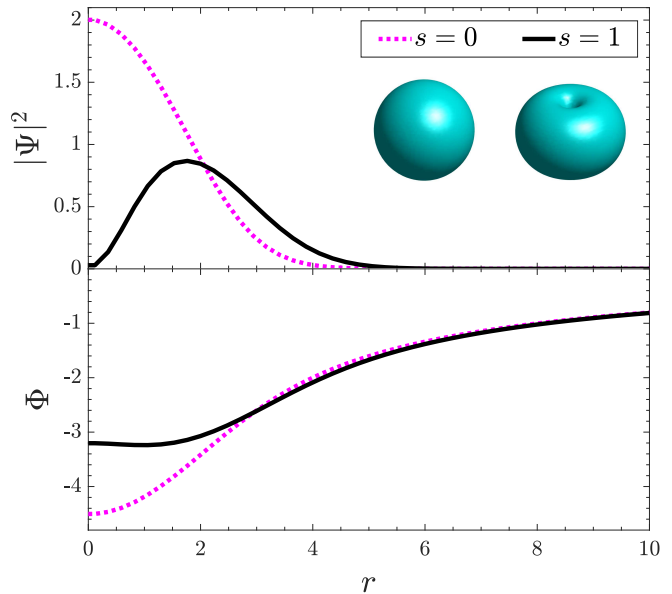}
\caption{Radial distribution  of the stationary density distribution and gravitational potential in $z=0$ plane for the BEC states $s=0$ and $s=1$ with normalization constant $N_{\text{s}}=100$. The inset represents the corresponding 3D isosurfaces of  condensate density.}
\label{fig:stst}
\end{center}
\end{figure}

Figure \ref{fig:stst} shows the density distribution $|\Psi|^2$ and the corresponding gravitational potential $\Phi$ for solutions with $s=0$ and $s=1$. As can be seen from the isosurfaces, the fundamental soliton is spherically symmetric, while the vortex has an axial symmetry. The peak density of the $s=0$ state is higher than in the case of the vortex soliton, whose density is more spread and decreases to zero at the origin. The gravitational potential at this point has the global minimum in the case $s=0$ and a local maximum in the case $s=1$.

\begin{figure}[htb]
\begin{center}
\includegraphics[width=3.4in]{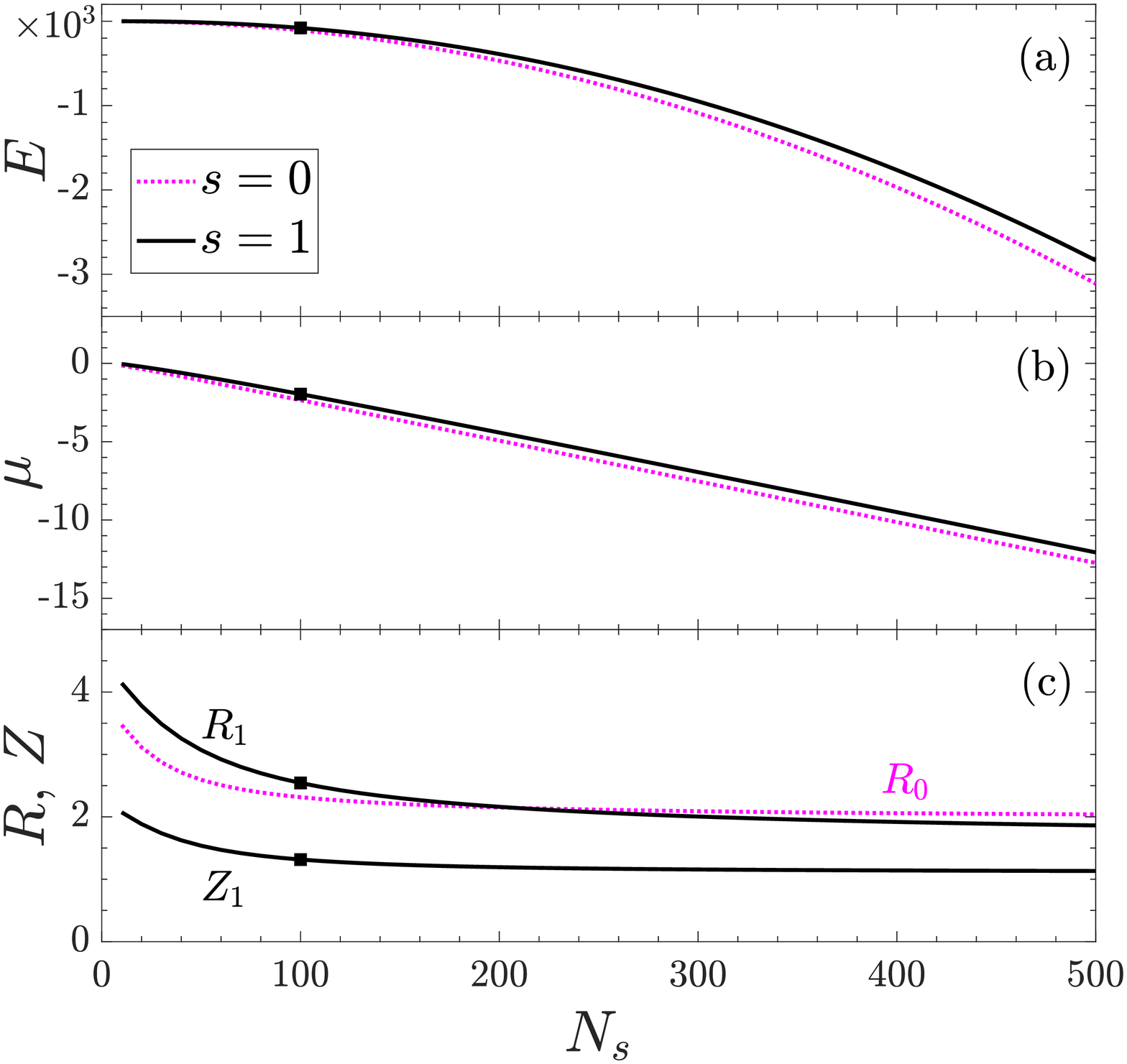}
\caption{(a) Total energy $E$, (b) the chemical potential $\mu$, (c) effective radius $R$, and height $Z$ as functions of the normalization constant $N_{\text{s}}$ for stationary states with topological charges $s=0$ and $s=1$. Black dots show the corresponding values at $N_{\text{s}}=100$.}
\label{fig:st_int}
\end{center}
\end{figure}

The parameters which characterize the system size can be defined as
\begin{equation}
R_0^2 = \frac{1}{N_{\text{s}}}\int r^2 |\Psi|^2 d\textbf{r}
    \label{eq:R_0}
\end{equation}
for the $s=0$ state, where $r = \sqrt{x^2+y^2+z^2}$, and
\begin{equation}
R_1^2 = \frac{1}{N_{\text{s}}}\int r_\perp^2 |\Psi|^2 d\textbf{r},
    \label{eq:R_1}
\end{equation}
\begin{equation}
Z_1^2  
= \frac{1}{N_{\text{s}}}\int z^2 |\Psi|^2 d\textbf{r}
\label{eq:Z_1}
\end{equation}
for the $s=1$ state. 
Figure \ref{fig:st_int} represents the total energy defined by Eq.(\ref{Etot}), the chemical potential, determined by Eq.(\ref{eq:stat_st}), and the system size characteristics given by Eqs.(\ref{eq:R_0})-(\ref{eq:Z_1}) as functions of normalization constant $N_{\text{s}}$.
As one can see, the energy and chemical potential of the vortex state are bigger due to additional rotational energy. The size of both systems decreases with the mass and approaches a constant value. Due to the vortex core structure, the mean width of the vortex state is approximately two times bigger than its height. The black dots mark the value corresponding to the considered normalization constant $N_{\text{s}}=100$.

As it has been mentioned above, the two considered self-gravitating BEC states with $s=0$ and $s=1$ are quite robust, which makes interesting studying their collisions.

\section{Binary collisions}
\label{sec:binary}

Here, we consider the scattering of the solitons discussed in the previous section. 
Initially, we place two BEC clouds of the same mass at some distance from each other so that their wave functions do not overlap. We set the initial velocities along the $z$-axis by multiplying the wave functions by the appropriate exponents $\Psi=\Psi_1 \cdot e^{-ivz}+\Psi_2 \cdot e^{ivz}$.
For sufficiently small initial velocities, condensate structures can merge and form a single blob. 
In all further cases, we consider initial dimensionless velocity $v=1.5$ (in units of $L_*\Omega_*$) large enough to produce quasi-elastic collisions.

\begin{figure*}[htp!]
\begin{center}
\includegraphics[width=6.8in]{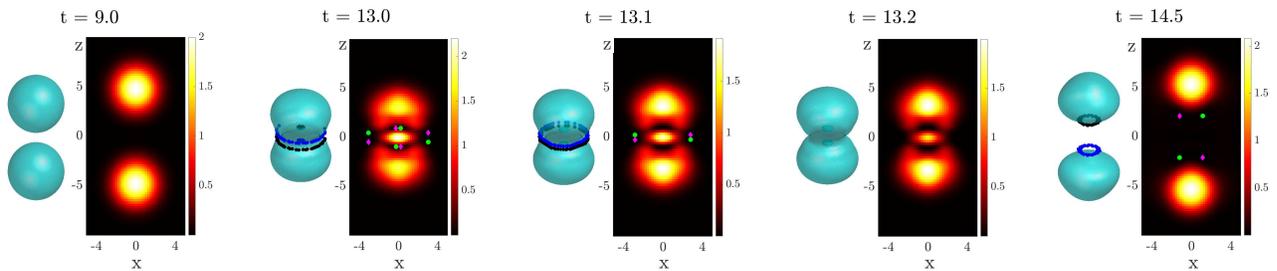}
\caption{A sequence of the density isosurfaces representing the dynamics of $s=0$ solitons collision with the formation of oppositely rotating vortex rings (colored in blue and black).  Snapshots show the density distribution in the plane $(y=0)$. Green dots represent the location of vortex cores, and magenta diamonds correspond to antivortex core location.}
\label{fig:pic_s=0}
\end{center}
\end{figure*}

Let us first investigate the collision of two fundamental solitons ($s=0$). During the merging, there appear such structures as vortex rings, which can be seen in Fig. \ref{fig:pic_s=0}. The cyan structures show the surfaces of the constant density value. The blue and black dots mark the centers of the vortex cores and represent oppositely rotating vortex rings.
As can be seen, when approaching, pairs of oppositely rotating vortex rings (marked in blue and black on the isosurfaces) are formed.
Further, the inner rings shrink to a point and disappear. The outer pair of rings annihilates as they approach. The rings are again visible when the solitons move away.

\begin{figure*}[ht]
\begin{center}
\includegraphics[width=6.8in]{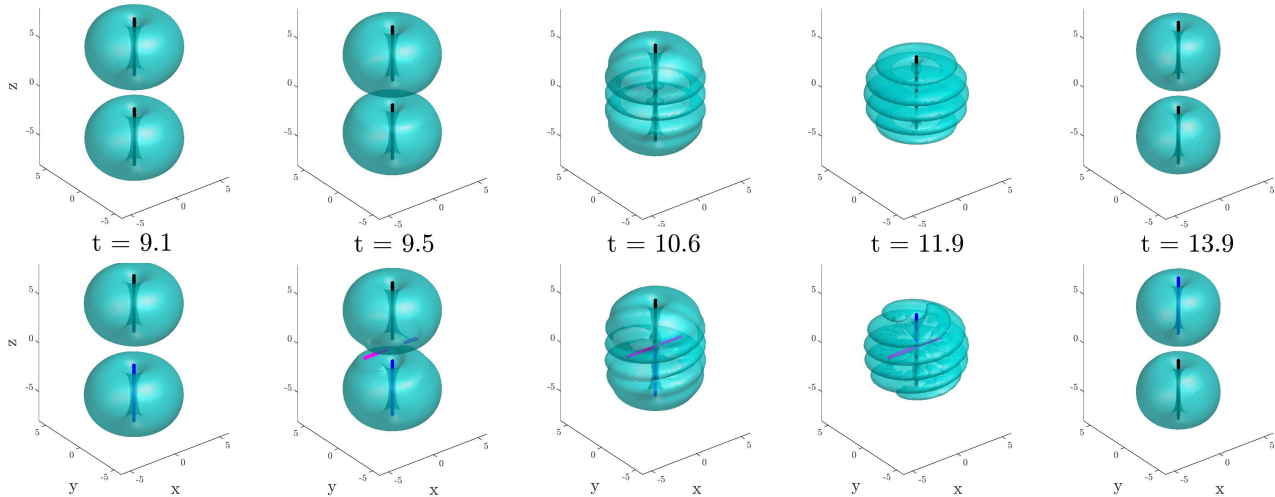}
\caption{Snapshots of head-on collision of BEC vortex states propagating along the vortex line direction. Shown are the condensate density isosurfaces for colliding vortex structures  with the co-rotating (upper row) and counter-rotating (lower row) superflows. The dots represent the vortex cores' location, and their colors correspond to the direction of the superflow rotation.}
\label{fig:PA_Coaxial}
\end{center}
\end{figure*}

Now let us consider collisions of single-charged vortex solitons ($s=1$).  
The patterns appearing during the dynamics depend on the orientation of the vortices' angular momenta.
Figure \ref{fig:PA_Coaxial} shows the simulation of the collision dynamics of the vortices with angular momenta parallel to the collision axis.
The upper row represents the case of the co-rotating superflows ($s_1=1$, $s_2=1$), while the lower row corresponds to the counter-rotating superflows ($s_1=1$, $s_2=-1$). Blue and black dots indicate the location of the vortices' cores and their orientation.
During merging, solitons interfere but pass through each other and revive their forms. 
As one can see in the case of oppositely oriented angular momenta, there appears a perpendicular vortex core (marked by magenta dots) similar to the well-known Josephson vortex (see e.g., \cite{2022PhRvA.106d3305B}) which is responsible for the continuity of the wave function phase.

\begin{figure*}[ht]
\begin{center}
\includegraphics[width=6.8in]{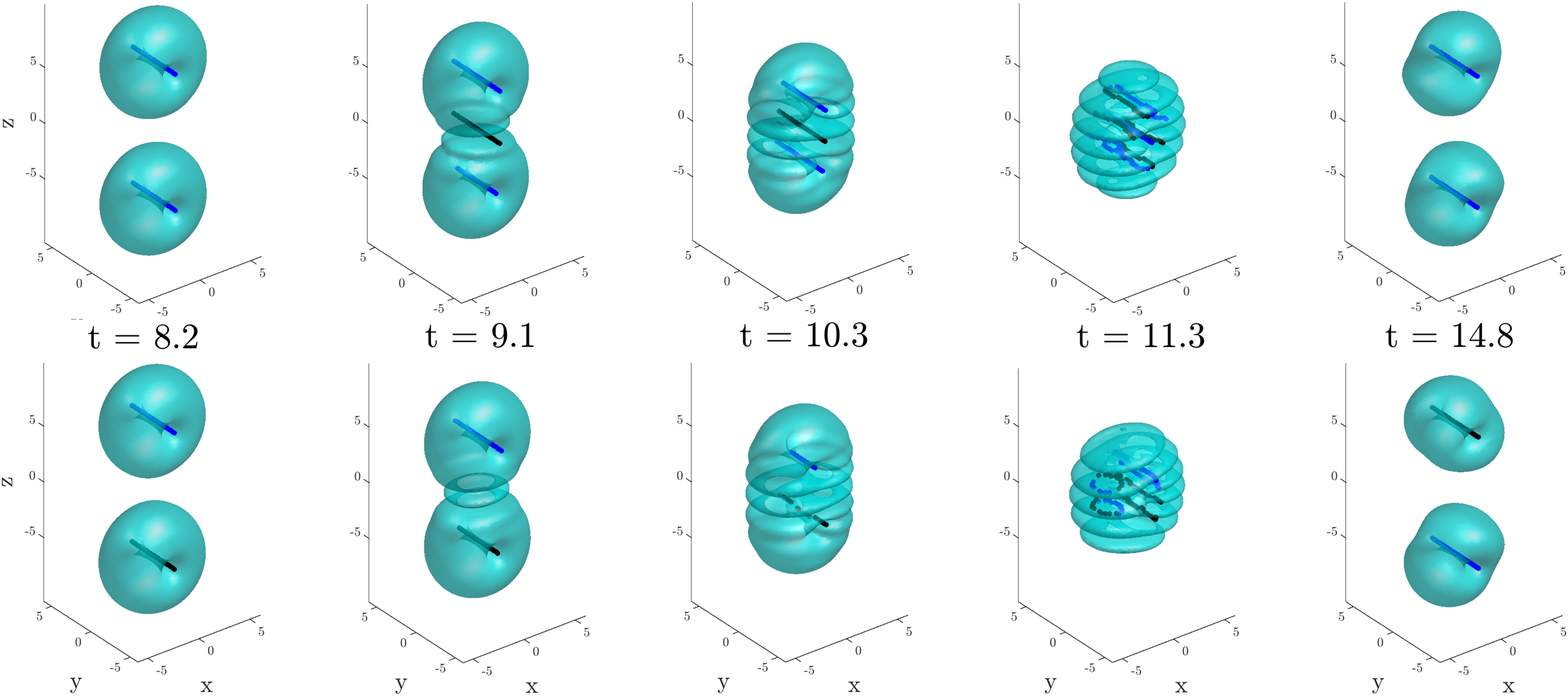}
\caption{The same as in Fig.\ref{fig:PA_Coaxial} but with the angular momentum perpendicular to the collision axis. The upper row shows the colliding of co-rotating superflows, and the lower row represents the case of counter-rotating superflows.}
\label{fig:PA_Coplanar}
\end{center}
\end{figure*}

Further, we investigate another limiting case of the vortex cores' orientation during the scattering: perpendicular to the axis of colliding. Figure \ref {fig:PA_Coplanar} shows a collision of the co-rotational ($s_1=1$, $s_2=1$) and counter-rotational superflows ($s_1=1$, $s_2=-1$), with angular momentum directed along the $y$-axis.
In this configuration, an additional vortex appears in the case of co-rotational superflows collision. 
One can also see the complicated dynamics with vortex lines' recombination into the vortex rings. And as in the previous case, vortices renew their structure after the merging.

\begin{figure*}[ht]
\begin{center}
\includegraphics[width=6.8in]{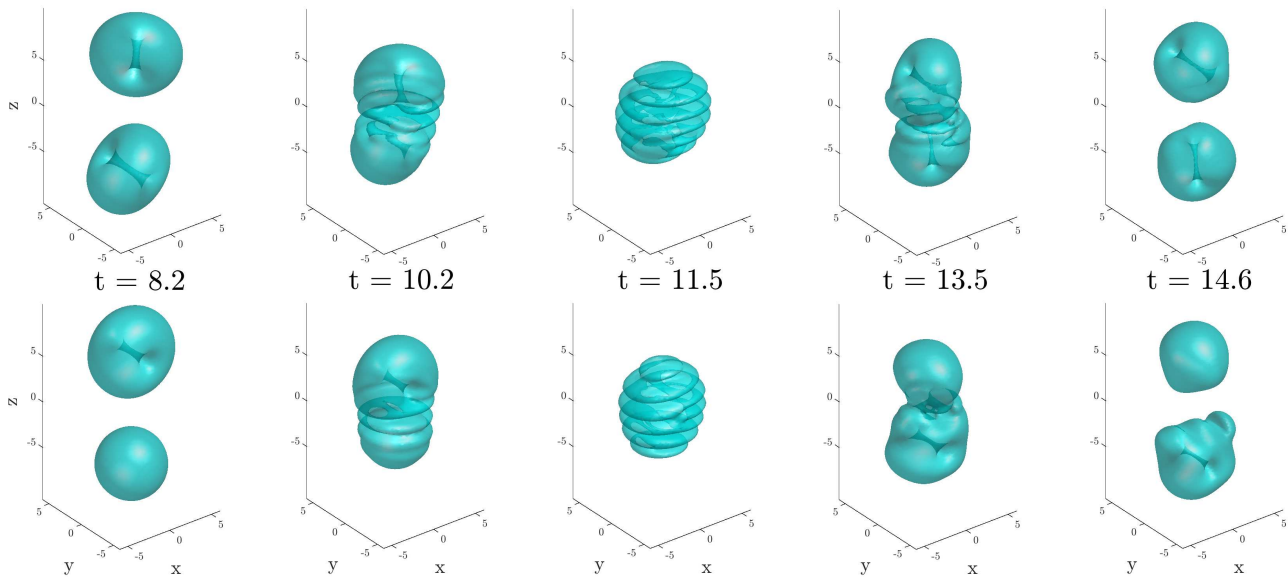}
\caption{The upper row demonstrates snapshots of the collision of two vortex states with arbitrary oriented angular momentum and non-zero impact parameter. The lower row shows the collision of the fundamental soliton and the vortex, with angular momentum perpendicular to the direction of movement.}
\label{fig:RandomlyIP_VortexSoliton}
\end{center}
\end{figure*}

In all previous cases of the symmetric superflows orientation, solitons have not been destroyed despite the density redistribution during the collision. Therefore, it is interesting to investigate the interaction of vortices with arbitrarily directed angular momenta. It turns out that vortices revive after merging, even in the general case of the arbitrary orientation and non-zero impact parameter (the upper row in Fig.\ref{fig:RandomlyIP_VortexSoliton}). 
We have also considered the collision of a vortex ($s_1=1$) and a fundamental soliton ($s_2=0$), which is shown on the lower row in Fig.\ref{fig:RandomlyIP_VortexSoliton}. As in all previous cases, the solitons pass through each other without being destroyed.
In these cases, we do not show the vortex cores, as the dynamic of collisions is complicated and governed by initial conditions.

\begin{figure}[htp!]
\begin{center}
\includegraphics[width=3.4in]{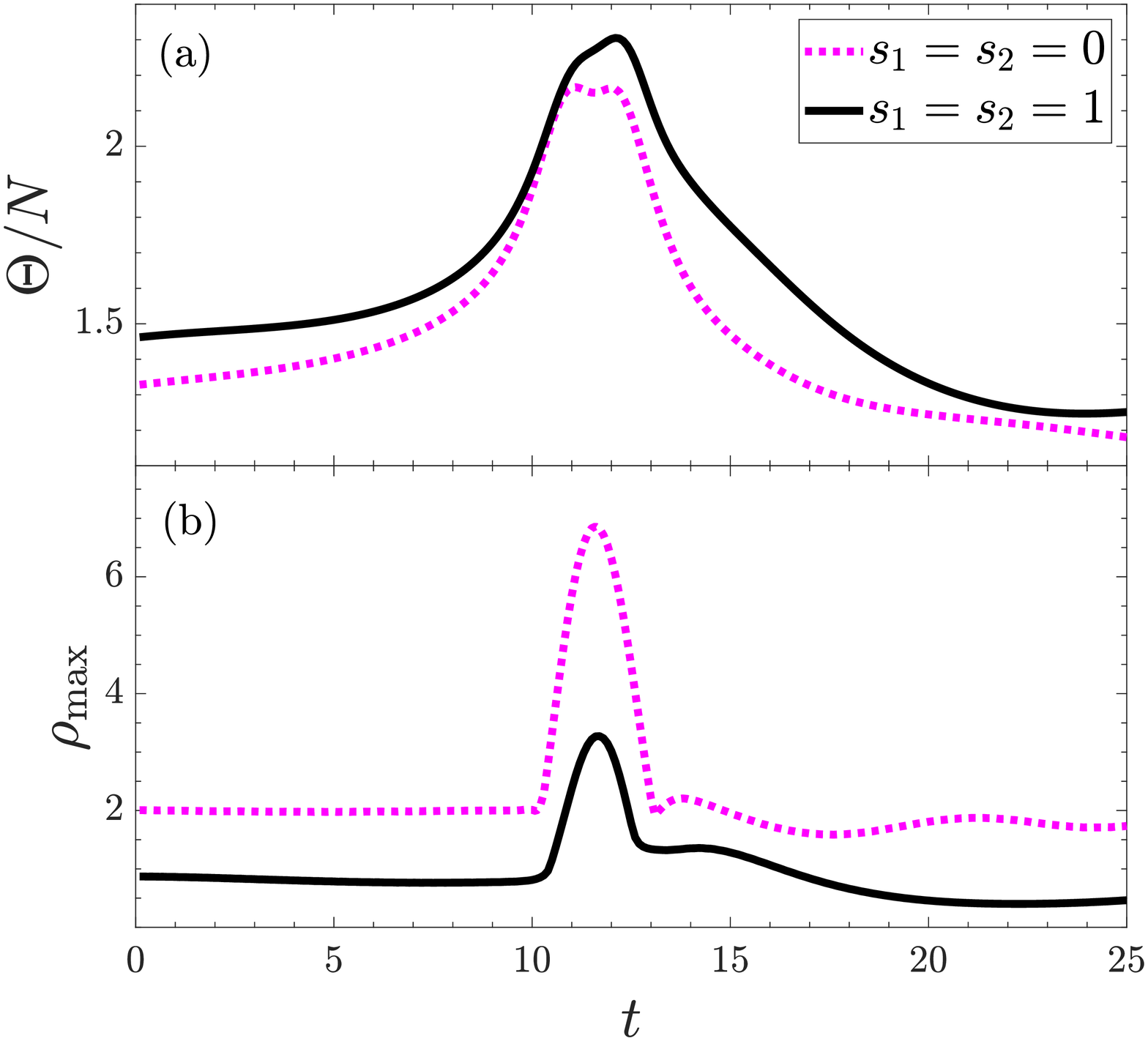}
\caption{ (a) Kinetic energy per particle, (b) peak density value  during the collision of two DM. The magenta line represents the case of colliding solitons with topological charges $s_1=s_2=0$, while the black line corresponds to $s_1=s_2=1$ vortices propagating along the direction of the angular momenta.}
\label{fig:s=01_int}
\end{center}
\end{figure}

Figure \ref{fig:s=01_int} shows the comparison between the merging dynamics of  fundamental solitons (magenta line) and vortices moving along the direction of the vortex line (black line). The upper panel represents the relative kinetic energy  $\Theta/N$, (where $N=2N_{\text{s}}$ the total norm of the system), which is higher in the case of vortex collision due to the nonzero angular momentum. Note that the final value of the kinetic energy is lower than the initial one, which illustrates that collision is not completely elastic.
The lower panel shows the peak density of the system as a function of time. As can be seen, in the case of merging $s=1$ vortex solitons, its variation is bigger.


In the present work, we have concentrated on collisions of DM structures with high enough relative velocities, which corresponds to the quasi-elastic interaction. We have also investigated the system with lower initial kinetic energy, which leads to the formation of the single localized structure after long-term evolution. The final state of the system is determined by the initial angular momenta. For example, merging vortices propagating along the vortex line with $s_1=1$ and $s_2=-1$ leads to the formation of fundamental soliton $s=0$, while in the case of $s_1=1$ and $s_2=1$ we can obtain $s=1$ vortex structure.

{In all the collisions mentioned above, especially during the merging of two fundamental solitons/vortices, we have seen complex redistribution of DM density. It is well known that such matter redistribution will generate GWs. 
With the fast development of GW detection techniques in recent years, such GWs, once observed, could become a new tool to investigate the properties of the relevant BEC DM. 
Therefore, in the next Section, we investigate the GW radiation in the collision processes studied above.}

\section{Gravitational wave radiation}
\label{sec:GW}



Since the velocity of the BEC matter studied here is non-relativistic, the energy-momentum tensor is dominated by the rest energy, i.e., the density distribution function. Consequently, we can directly use the density distribution found in Sec. \ref{sec:binary} to compute the GW luminosity.  
We will consider quadrupole moments and related GW radiation, and we will also use the common assumption that the source-detector distance is much larger than the source size \cite{poisson2014gravity}. 

With these considerations, then the luminosity of the GW takes the standard form \cite{quilis}
\begin{equation}
L_{\text{GW}}=C\sum_{i,j}|\dddot{I}_{ij}|^2,
\label{eq:gwlum}
\end{equation}
where $C=\frac{G^4}{80\pi^2 c^5}\frac{m^5}{ a_s^5}$ and $\dddot{I}_{ij}$ is the third derivative of the following dimensionless quadrupole moment defined in (\ref{eq:I_ij}) with respect to the dimensionless time $t$.


\begin{figure}[ht]
\begin{center}
\includegraphics[width=3.4in]{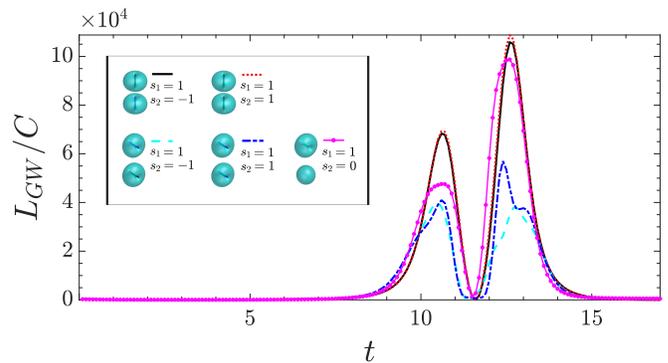}
\caption{The dimensionless luminosity of the gravity waves for different orientations of colliding the vortex structures shown in Figs. \ref{fig:PA_Coaxial}-\ref{fig:RandomlyIP_VortexSoliton}. Note that the luminosity of gravity waves is essentially affected by orientation of the superflows in the DM structures.
}
\label{fig:L_GW}
\end{center}
\end{figure}

The scaled GW luminosity $L_{\text{GW}}/C$ for collisions studied in Sec. \ref{sec:binary} is shown in Fig. \ref{fig:L_GW}.
It is seen that in all cases, the GW emission mainly happens
when the wave functions of the colliding DM structures substantially overlap. Note that the luminosity for the considered scattering processes is suppressed when the condensate density reaches its maximum value (see Fig. \ref{fig:s=01_int}) so that $L_{\text{GW}}$ forms a two-peak pattern. 
%
It is remarkable that the collisions of vortices with angular momenta along $z$-axis produce stronger emissions of GW.

Note that for the typical DM halo mass $M\sim 10^{12}M_\odot$ \cite{quilis} and boson particle mass of order $m\sim10^{-22}~\text{[eV]}$ the peak value of the GW luminosity,  max$L_{\text{GW}}\approx 2.7\times 10^{31}$ [erg/s], obtained for colliding superfluid DM structures is consistent with the results of Refs.  \cite{Inagaki:2010nu,quilis}. GWs of such luminosity and frequency are believed to have potential observable effects on the cosmic microwave background polarization \cite{quilis}. 




\section{Conclusions}
\label{sec:conclusions}
We have investigated the general properties and stability of solitonic structures of self-gravitating Bose-Einstein condensates. Previously, it was found that single fundamental and  one-charged vortex solitons are stable with respect to strong perturbations. In the present work, we reveal that vortex solitons survive even after head-on collision. 
In spite of the non-elasticity of interaction, vortex solitons demonstrate robust evolution and restore their forms after collisions. We have found that the quantum nature of the self-gravitating Bose-Einstein condensates leads to the formation of remarkable topological excitations in the form of vortex rings and Josephson vortices during the collision of the DM solitonic structures. We have investigated  gravitational waves radiated by interacting DM halos and demonstrated the effect of the quantum interference patterns formed by colliding BECs on gravitational wave luminosity.

The fascinating physics of superfluid DM      leads to a number of observational signatures that might help to elucidate the basic properties of dark matter particles. In particular, the existence of stable DM vortex structures can significantly affect the dynamics of luminous matter, especially near the vortex core.    Detailed consideration of interactions between stable DM vortex structures and the baryonic matter may be a relevant extension of the present work. 


\section{Acknowledgment}
We thank Stanislav Vilchinskii and Luca Salasnich for useful discussions. A.Y. acknowledge support from BIRD Project "Ultracold atoms in curved geometries" of the University of Padova.

\bibliography{ref1}

\begin{thebibliography}{58}%
\makeatletter
\providecommand \@ifxundefined [1]{%
 \@ifx{#1\undefined}
}%
\providecommand \@ifnum [1]{%
 \ifnum #1\expandafter \@firstoftwo
 \else \expandafter \@secondoftwo
 \fi
}%
\providecommand \@ifx [1]{%
 \ifx #1\expandafter \@firstoftwo
 \else \expandafter \@secondoftwo
 \fi
}%
\providecommand \natexlab [1]{#1}%
\providecommand \enquote  [1]{``#1''}%
\providecommand \bibnamefont  [1]{#1}%
\providecommand \bibfnamefont [1]{#1}%
\providecommand \citenamefont [1]{#1}%
\providecommand \href@noop [0]{\@secondoftwo}%
\providecommand \href [0]{\begingroup \@sanitize@url \@href}%
\providecommand \@href[1]{\@@startlink{#1}\@@href}%
\providecommand \@@href[1]{\endgroup#1\@@endlink}%
\providecommand \@sanitize@url [0]{\catcode `\\12\catcode `\$12\catcode
  `\&12\catcode `\#12\catcode `\^12\catcode `\_12\catcode `\%12\relax}%
\providecommand \@@startlink[1]{}%
\providecommand \@@endlink[0]{}%
\providecommand \url  [0]{\begingroup\@sanitize@url \@url }%
\providecommand \@url [1]{\endgroup\@href {#1}{\urlprefix }}%
\providecommand \urlprefix  [0]{URL }%
\providecommand \Eprint [0]{\href }%
\providecommand \doibase [0]{https://doi.org/}%
\providecommand \selectlanguage [0]{\@gobble}%
\providecommand \bibinfo  [0]{\@secondoftwo}%
\providecommand \bibfield  [0]{\@secondoftwo}%
\providecommand \translation [1]{[#1]}%
\providecommand \BibitemOpen [0]{}%
\providecommand \bibitemStop [0]{}%
\providecommand \bibitemNoStop [0]{.\EOS\space}%
\providecommand \EOS [0]{\spacefactor3000\relax}%
\providecommand \BibitemShut  [1]{\csname bibitem#1\endcsname}%
\let\auto@bib@innerbib\@empty
\bibitem [{\citenamefont {{Edmonds}}\ \emph {et~al.}(2017)\citenamefont
  {{Edmonds}}, \citenamefont {{Bland}}, \citenamefont {{Doran}},\ and\
  \citenamefont {{Parker}}}]{2017NJPh...19b3019E}%
  \BibitemOpen
  \bibfield  {author} {\bibinfo {author} {\bibfnamefont {M.~J.}\ \bibnamefont
  {{Edmonds}}}, \bibinfo {author} {\bibfnamefont {T.}~\bibnamefont {{Bland}}},
  \bibinfo {author} {\bibfnamefont {R.}~\bibnamefont {{Doran}}},\ and\ \bibinfo
  {author} {\bibfnamefont {N.~G.}\ \bibnamefont {{Parker}}},\ }\href
  {https://doi.org/10.1088/1367-2630/aa5a6b} {\bibfield  {journal} {\bibinfo
  {journal} {New Journal of Physics}\ }\textbf {\bibinfo {volume} {19}},\
  \bibinfo {eid} {023019} (\bibinfo {year} {2017})},\ \Eprint
  {https://arxiv.org/abs/1610.01022} {arXiv:1610.01022 [cond-mat.quant-gas]}
  \BibitemShut {NoStop}%
\bibitem [{\citenamefont {{Schive}}\ \emph
  {et~al.}(2014{\natexlab{a}})\citenamefont {{Schive}}, \citenamefont
  {{Chiueh}},\ and\ \citenamefont {{Broadhurst}}}]{2014NatPh..10..496S}%
  \BibitemOpen
  \bibfield  {author} {\bibinfo {author} {\bibfnamefont {H.-Y.}\ \bibnamefont
  {{Schive}}}, \bibinfo {author} {\bibfnamefont {T.}~\bibnamefont {{Chiueh}}},\
  and\ \bibinfo {author} {\bibfnamefont {T.}~\bibnamefont {{Broadhurst}}},\
  }\href {https://doi.org/10.1038/nphys2996} {\bibfield  {journal} {\bibinfo
  {journal} {Nature Physics}\ }\textbf {\bibinfo {volume} {10}},\ \bibinfo
  {pages} {496} (\bibinfo {year} {2014}{\natexlab{a}})},\ \Eprint
  {https://arxiv.org/abs/1406.6586} {arXiv:1406.6586 [astro-ph.GA]}
  \BibitemShut {NoStop}%
\bibitem [{\citenamefont {{Matos}}\ and\ \citenamefont {{Arturo
  Ure{\~n}a-L{\'o}pez}}(2001)}]{2001PhRvD..63f3506M}%
  \BibitemOpen
  \bibfield  {author} {\bibinfo {author} {\bibfnamefont {T.}~\bibnamefont
  {{Matos}}}\ and\ \bibinfo {author} {\bibfnamefont {L.}~\bibnamefont {{Arturo
  Ure{\~n}a-L{\'o}pez}}},\ }\href {https://doi.org/10.1103/PhysRevD.63.063506}
  {\bibfield  {journal} {\bibinfo  {journal} {\prd}\ }\textbf {\bibinfo
  {volume} {63}},\ \bibinfo {eid} {063506} (\bibinfo {year} {2001})},\ \Eprint
  {https://arxiv.org/abs/astro-ph/0006024} {arXiv:astro-ph/0006024 [astro-ph]}
  \BibitemShut {NoStop}%
\bibitem [{\citenamefont {{Sahni}}\ and\ \citenamefont
  {{Wang}}(2000)}]{2000PhRvD..62j3517S}%
  \BibitemOpen
  \bibfield  {author} {\bibinfo {author} {\bibfnamefont {V.}~\bibnamefont
  {{Sahni}}}\ and\ \bibinfo {author} {\bibfnamefont {L.}~\bibnamefont
  {{Wang}}},\ }\href {https://doi.org/10.1103/PhysRevD.62.103517} {\bibfield
  {journal} {\bibinfo  {journal} {\prd}\ }\textbf {\bibinfo {volume} {62}},\
  \bibinfo {eid} {103517} (\bibinfo {year} {2000})},\ \Eprint
  {https://arxiv.org/abs/astro-ph/9910097} {arXiv:astro-ph/9910097 [astro-ph]}
  \BibitemShut {NoStop}%
\bibitem [{\citenamefont {Goldstein}\ \emph {et~al.}(2022)\citenamefont
  {Goldstein}, \citenamefont {Koushiappas},\ and\ \citenamefont
  {Walker}}]{Goldstein_2022}%
  \BibitemOpen
  \bibfield  {author} {\bibinfo {author} {\bibfnamefont {I.~S.}\ \bibnamefont
  {Goldstein}}, \bibinfo {author} {\bibfnamefont {S.~M.}\ \bibnamefont
  {Koushiappas}},\ and\ \bibinfo {author} {\bibfnamefont {M.~G.}\ \bibnamefont
  {Walker}},\ }\bibfield  {journal} {\bibinfo  {journal} {Physical Review D}\
  }\textbf {\bibinfo {volume} {106}},\ \href
  {https://doi.org/10.1103/physrevd.106.063010} {10.1103/physrevd.106.063010}
  (\bibinfo {year} {2022})\BibitemShut {NoStop}%
\bibitem [{\citenamefont {{Paredes}}\ \emph {et~al.}(2020)\citenamefont
  {{Paredes}}, \citenamefont {{Olivieri}},\ and\ \citenamefont
  {{Michinel}}}]{2020PhyD..40332301P}%
  \BibitemOpen
  \bibfield  {author} {\bibinfo {author} {\bibfnamefont {A.}~\bibnamefont
  {{Paredes}}}, \bibinfo {author} {\bibfnamefont {D.~N.}\ \bibnamefont
  {{Olivieri}}},\ and\ \bibinfo {author} {\bibfnamefont {H.}~\bibnamefont
  {{Michinel}}},\ }\href {https://doi.org/10.1016/j.physd.2019.132301}
  {\bibfield  {journal} {\bibinfo  {journal} {Physica D Nonlinear Phenomena}\
  }\textbf {\bibinfo {volume} {403}},\ \bibinfo {eid} {132301} (\bibinfo {year}
  {2020})}\BibitemShut {NoStop}%
\bibitem [{\citenamefont {Maucher}\ \emph {et~al.}(2010)\citenamefont
  {Maucher}, \citenamefont {Skupin}, \citenamefont {Shen},\ and\ \citenamefont
  {Krolikowski}}]{PhysRevA.81.063617}%
  \BibitemOpen
  \bibfield  {author} {\bibinfo {author} {\bibfnamefont {F.}~\bibnamefont
  {Maucher}}, \bibinfo {author} {\bibfnamefont {S.}~\bibnamefont {Skupin}},
  \bibinfo {author} {\bibfnamefont {M.}~\bibnamefont {Shen}},\ and\ \bibinfo
  {author} {\bibfnamefont {W.}~\bibnamefont {Krolikowski}},\ }\href
  {https://doi.org/10.1103/PhysRevA.81.063617} {\bibfield  {journal} {\bibinfo
  {journal} {Phys. Rev. A}\ }\textbf {\bibinfo {volume} {81}},\ \bibinfo
  {pages} {063617} (\bibinfo {year} {2010})}\BibitemShut {NoStop}%
\bibitem [{\citenamefont {O'Dell}\ \emph {et~al.}(2000)\citenamefont {O'Dell},
  \citenamefont {Giovanazzi}, \citenamefont {Kurizki},\ and\ \citenamefont
  {Akulin}}]{PhysRevLett.84.5687}%
  \BibitemOpen
  \bibfield  {author} {\bibinfo {author} {\bibfnamefont {D.}~\bibnamefont
  {O'Dell}}, \bibinfo {author} {\bibfnamefont {S.}~\bibnamefont {Giovanazzi}},
  \bibinfo {author} {\bibfnamefont {G.}~\bibnamefont {Kurizki}},\ and\ \bibinfo
  {author} {\bibfnamefont {V.~M.}\ \bibnamefont {Akulin}},\ }\href
  {https://doi.org/10.1103/PhysRevLett.84.5687} {\bibfield  {journal} {\bibinfo
   {journal} {Phys. Rev. Lett.}\ }\textbf {\bibinfo {volume} {84}},\ \bibinfo
  {pages} {5687} (\bibinfo {year} {2000})}\BibitemShut {NoStop}%
\bibitem [{\citenamefont {Giovanazzi}\ \emph {et~al.}(2001)\citenamefont
  {Giovanazzi}, \citenamefont {O'Dell},\ and\ \citenamefont
  {Kurizki}}]{PhysRevA.63.031603}%
  \BibitemOpen
  \bibfield  {author} {\bibinfo {author} {\bibfnamefont {S.}~\bibnamefont
  {Giovanazzi}}, \bibinfo {author} {\bibfnamefont {D.}~\bibnamefont {O'Dell}},\
  and\ \bibinfo {author} {\bibfnamefont {G.}~\bibnamefont {Kurizki}},\ }\href
  {https://doi.org/10.1103/PhysRevA.63.031603} {\bibfield  {journal} {\bibinfo
  {journal} {Phys. Rev. A}\ }\textbf {\bibinfo {volume} {63}},\ \bibinfo
  {pages} {031603} (\bibinfo {year} {2001})}\BibitemShut {NoStop}%
\bibitem [{\citenamefont {{Giovanazzi}}\ \emph {et~al.}(2001)\citenamefont
  {{Giovanazzi}}, \citenamefont {{Kurizki}}, \citenamefont {{Mazets}},\ and\
  \citenamefont {{Stringari}}}]{2001EL.....56....1G}%
  \BibitemOpen
  \bibfield  {author} {\bibinfo {author} {\bibfnamefont {S.}~\bibnamefont
  {{Giovanazzi}}}, \bibinfo {author} {\bibfnamefont {G.}~\bibnamefont
  {{Kurizki}}}, \bibinfo {author} {\bibfnamefont {I.~E.}\ \bibnamefont
  {{Mazets}}},\ and\ \bibinfo {author} {\bibfnamefont {S.}~\bibnamefont
  {{Stringari}}},\ }\href {https://doi.org/10.1209/epl/i2001-00478-8}
  {\bibfield  {journal} {\bibinfo  {journal} {EPL (Europhysics Letters)}\
  }\textbf {\bibinfo {volume} {56}},\ \bibinfo {pages} {1} (\bibinfo {year}
  {2001})},\ \Eprint {https://arxiv.org/abs/cond-mat/0101310}
  {arXiv:cond-mat/0101310 [cond-mat]} \BibitemShut {NoStop}%
\bibitem [{\citenamefont {Böhmer}\ and\ \citenamefont
  {Harko}(2007)}]{2007JCAP...06..025B}%
  \BibitemOpen
  \bibfield  {author} {\bibinfo {author} {\bibfnamefont {C.~G.}\ \bibnamefont
  {Böhmer}}\ and\ \bibinfo {author} {\bibfnamefont {T.}~\bibnamefont
  {Harko}},\ }\href {https://doi.org/10.1088/1475-7516/2007/06/025} {\bibfield
  {journal} {\bibinfo  {journal} {Journal of Cosmology and Astroparticle
  Physics}\ }\textbf {\bibinfo {volume} {2007}}\bibinfo  {number} { (06)},\
  \bibinfo {pages} {025}}\BibitemShut {NoStop}%
\bibitem [{\citenamefont {{Chavanis}}(2017)}]{2017EPJP..132..248C}%
  \BibitemOpen
\bibfield  {number} {  }\bibfield  {author} {\bibinfo {author} {\bibfnamefont
  {P.-H.}\ \bibnamefont {{Chavanis}}},\ }\href
  {https://doi.org/10.1140/epjp/i2017-11544-3} {\bibfield  {journal} {\bibinfo
  {journal} {European Physical Journal Plus}\ }\textbf {\bibinfo {volume}
  {132}},\ \bibinfo {eid} {248} (\bibinfo {year} {2017})},\ \Eprint
  {https://arxiv.org/abs/1611.09610} {arXiv:1611.09610 [gr-qc]} \BibitemShut
  {NoStop}%
\bibitem [{\citenamefont {Chavanis}\ and\ \citenamefont
  {Harko}(2012)}]{Chavanis_2012}%
  \BibitemOpen
  \bibfield  {author} {\bibinfo {author} {\bibfnamefont {P.-H.}\ \bibnamefont
  {Chavanis}}\ and\ \bibinfo {author} {\bibfnamefont {T.}~\bibnamefont
  {Harko}},\ }\bibfield  {journal} {\bibinfo  {journal} {Physical Review D}\
  }\textbf {\bibinfo {volume} {86}},\ \href
  {https://doi.org/10.1103/physrevd.86.064011} {10.1103/physrevd.86.064011}
  (\bibinfo {year} {2012})\BibitemShut {NoStop}%
\bibitem [{\citenamefont {Hui}\ \emph {et~al.}(2017)\citenamefont {Hui},
  \citenamefont {Ostriker}, \citenamefont {Tremaine},\ and\ \citenamefont
  {Witten}}]{Hui_2017}%
  \BibitemOpen
  \bibfield  {author} {\bibinfo {author} {\bibfnamefont {L.}~\bibnamefont
  {Hui}}, \bibinfo {author} {\bibfnamefont {J.~P.}\ \bibnamefont {Ostriker}},
  \bibinfo {author} {\bibfnamefont {S.}~\bibnamefont {Tremaine}},\ and\
  \bibinfo {author} {\bibfnamefont {E.}~\bibnamefont {Witten}},\ }\bibfield
  {journal} {\bibinfo  {journal} {Physical Review D}\ }\textbf {\bibinfo
  {volume} {95}},\ \href {https://doi.org/10.1103/physrevd.95.043541}
  {10.1103/physrevd.95.043541} (\bibinfo {year} {2017})\BibitemShut {NoStop}%
\bibitem [{\citenamefont {{Madarassy}}\ and\ \citenamefont
  {{Toth}}(2015)}]{2015PhRvD..91d4041M}%
  \BibitemOpen
  \bibfield  {author} {\bibinfo {author} {\bibfnamefont {E.~J.~M.}\
  \bibnamefont {{Madarassy}}}\ and\ \bibinfo {author} {\bibfnamefont {V.~T.}\
  \bibnamefont {{Toth}}},\ }\href {https://doi.org/10.1103/PhysRevD.91.044041}
  {\bibfield  {journal} {\bibinfo  {journal} {\prd}\ }\textbf {\bibinfo
  {volume} {91}},\ \bibinfo {eid} {044041} (\bibinfo {year} {2015})},\ \Eprint
  {https://arxiv.org/abs/1412.7152} {arXiv:1412.7152 [hep-ph]} \BibitemShut
  {NoStop}%
\bibitem [{\citenamefont {Siemonsen}\ and\ \citenamefont
  {East}(2021)}]{Siemonsen_2021}%
  \BibitemOpen
  \bibfield  {author} {\bibinfo {author} {\bibfnamefont {N.}~\bibnamefont
  {Siemonsen}}\ and\ \bibinfo {author} {\bibfnamefont {W.~E.}\ \bibnamefont
  {East}},\ }\bibfield  {journal} {\bibinfo  {journal} {Physical Review D}\
  }\textbf {\bibinfo {volume} {103}},\ \href
  {https://doi.org/10.1103/physrevd.103.044022} {10.1103/physrevd.103.044022}
  (\bibinfo {year} {2021})\BibitemShut {NoStop}%
\bibitem [{\citenamefont {Yakimenko}\ \emph {et~al.}(2005)\citenamefont
  {Yakimenko}, \citenamefont {Zaliznyak},\ and\ \citenamefont
  {Kivshar}}]{yakimenko2005stable}%
  \BibitemOpen
  \bibfield  {author} {\bibinfo {author} {\bibfnamefont {A.~I.}\ \bibnamefont
  {Yakimenko}}, \bibinfo {author} {\bibfnamefont {Y.~A.}\ \bibnamefont
  {Zaliznyak}},\ and\ \bibinfo {author} {\bibfnamefont {Y.}~\bibnamefont
  {Kivshar}},\ }\href@noop {} {\bibfield  {journal} {\bibinfo  {journal}
  {Physical Review E}\ }\textbf {\bibinfo {volume} {71}},\ \bibinfo {pages}
  {065603} (\bibinfo {year} {2005})}\BibitemShut {NoStop}%
\bibitem [{\citenamefont {Lashkin}\ \emph {et~al.}(2009)\citenamefont
  {Lashkin}, \citenamefont {Yakimenko},\ and\ \citenamefont
  {Zaliznyak}}]{lashkin2009stable}%
  \BibitemOpen
  \bibfield  {author} {\bibinfo {author} {\bibfnamefont {V.}~\bibnamefont
  {Lashkin}}, \bibinfo {author} {\bibfnamefont {A.}~\bibnamefont {Yakimenko}},\
  and\ \bibinfo {author} {\bibfnamefont {Y.~A.}\ \bibnamefont {Zaliznyak}},\
  }\href@noop {} {\bibfield  {journal} {\bibinfo  {journal} {Physica Scripta}\
  }\textbf {\bibinfo {volume} {79}},\ \bibinfo {pages} {035305} (\bibinfo
  {year} {2009})}\BibitemShut {NoStop}%
\bibitem [{\citenamefont {Rindler-Daller}\ and\ \citenamefont
  {Shapiro}(2012)}]{Rindler_Daller_2012}%
  \BibitemOpen
  \bibfield  {author} {\bibinfo {author} {\bibfnamefont {T.}~\bibnamefont
  {Rindler-Daller}}\ and\ \bibinfo {author} {\bibfnamefont {P.~R.}\
  \bibnamefont {Shapiro}},\ }\href
  {https://doi.org/10.1111/j.1365-2966.2012.20588.x} {\bibfield  {journal}
  {\bibinfo  {journal} {Monthly Notices of the Royal Astronomical Society}\
  }\textbf {\bibinfo {volume} {422}},\ \bibinfo {pages} {135} (\bibinfo {year}
  {2012})}\BibitemShut {NoStop}%
\bibitem [{\citenamefont {Sanchis-Gual}\ \emph {et~al.}(2019)\citenamefont
  {Sanchis-Gual}, \citenamefont {Giovanni}, \citenamefont {Zilh{\~{a} }o},
  \citenamefont {Herdeiro}, \citenamefont {Cerd{\'{a}}-Dur{\'{a}}n},
  \citenamefont {Font},\ and\ \citenamefont {Radu}}]{Sanchis_Gual_2019}%
  \BibitemOpen
  \bibfield  {author} {\bibinfo {author} {\bibfnamefont {N.}~\bibnamefont
  {Sanchis-Gual}}, \bibinfo {author} {\bibfnamefont {F.~D.}\ \bibnamefont
  {Giovanni}}, \bibinfo {author} {\bibfnamefont {M.}~\bibnamefont {Zilh{\~{a}
  }o}}, \bibinfo {author} {\bibfnamefont {C.}~\bibnamefont {Herdeiro}},
  \bibinfo {author} {\bibfnamefont {P.}~\bibnamefont
  {Cerd{\'{a}}-Dur{\'{a}}n}}, \bibinfo {author} {\bibfnamefont
  {J.}~\bibnamefont {Font}},\ and\ \bibinfo {author} {\bibfnamefont
  {E.}~\bibnamefont {Radu}},\ }\bibfield  {journal} {\bibinfo  {journal}
  {Physical Review Letters}\ }\textbf {\bibinfo {volume} {123}},\ \href
  {https://doi.org/10.1103/physrevlett.123.221101}
  {10.1103/physrevlett.123.221101} (\bibinfo {year} {2019})\BibitemShut
  {NoStop}%
\bibitem [{\citenamefont {Zhang}\ \emph {et~al.}(2018)\citenamefont {Zhang},
  \citenamefont {Chan}, \citenamefont {Harko}, \citenamefont {Liang},\ and\
  \citenamefont {Leung}}]{zhang2018slowly}%
  \BibitemOpen
  \bibfield  {author} {\bibinfo {author} {\bibfnamefont {X.}~\bibnamefont
  {Zhang}}, \bibinfo {author} {\bibfnamefont {M.~H.}\ \bibnamefont {Chan}},
  \bibinfo {author} {\bibfnamefont {T.}~\bibnamefont {Harko}}, \bibinfo
  {author} {\bibfnamefont {S.-D.}\ \bibnamefont {Liang}},\ and\ \bibinfo
  {author} {\bibfnamefont {C.~S.}\ \bibnamefont {Leung}},\ }\href@noop {}
  {\bibfield  {journal} {\bibinfo  {journal} {The European Physical Journal C}\
  }\textbf {\bibinfo {volume} {78}},\ \bibinfo {pages} {1} (\bibinfo {year}
  {2018})}\BibitemShut {NoStop}%
\bibitem [{\citenamefont {{Nikolaieva}}\ \emph {et~al.}(2021)\citenamefont
  {{Nikolaieva}}, \citenamefont {{Olashyn}}, \citenamefont {{Kuriatnikov}},
  \citenamefont {{Vilchynskii}},\ and\ \citenamefont
  {{Yakimenko}}}]{2021LTP....47..684N}%
  \BibitemOpen
  \bibfield  {author} {\bibinfo {author} {\bibfnamefont {Y.~O.}\ \bibnamefont
  {{Nikolaieva}}}, \bibinfo {author} {\bibfnamefont {A.~O.}\ \bibnamefont
  {{Olashyn}}}, \bibinfo {author} {\bibfnamefont {Y.~I.}\ \bibnamefont
  {{Kuriatnikov}}}, \bibinfo {author} {\bibfnamefont {S.~I.}\ \bibnamefont
  {{Vilchynskii}}},\ and\ \bibinfo {author} {\bibfnamefont {A.~I.}\
  \bibnamefont {{Yakimenko}}},\ }\href {https://doi.org/10.1063/10.0005557}
  {\bibfield  {journal} {\bibinfo  {journal} {Low Temperature Physics}\
  }\textbf {\bibinfo {volume} {47}},\ \bibinfo {pages} {684} (\bibinfo {year}
  {2021})},\ \Eprint {https://arxiv.org/abs/2103.07856} {arXiv:2103.07856
  [nlin.PS]} \BibitemShut {NoStop}%
\bibitem [{\citenamefont {Dmitriev}\ \emph {et~al.}(2021)\citenamefont
  {Dmitriev}, \citenamefont {Levkov}, \citenamefont {Panin}, \citenamefont
  {Pushnaya},\ and\ \citenamefont {Tkachev}}]{PhysRevD.104.023504}%
  \BibitemOpen
  \bibfield  {author} {\bibinfo {author} {\bibfnamefont {A.~S.}\ \bibnamefont
  {Dmitriev}}, \bibinfo {author} {\bibfnamefont {D.~G.}\ \bibnamefont
  {Levkov}}, \bibinfo {author} {\bibfnamefont {A.~G.}\ \bibnamefont {Panin}},
  \bibinfo {author} {\bibfnamefont {E.~K.}\ \bibnamefont {Pushnaya}},\ and\
  \bibinfo {author} {\bibfnamefont {I.~I.}\ \bibnamefont {Tkachev}},\ }\href
  {https://doi.org/10.1103/PhysRevD.104.023504} {\bibfield  {journal} {\bibinfo
   {journal} {Phys. Rev. D}\ }\textbf {\bibinfo {volume} {104}},\ \bibinfo
  {pages} {023504} (\bibinfo {year} {2021})}\BibitemShut {NoStop}%
\bibitem [{\citenamefont {Choi}(2002)}]{PhysRevA.66.063609}%
  \BibitemOpen
  \bibfield  {author} {\bibinfo {author} {\bibfnamefont {D.-I.}\ \bibnamefont
  {Choi}},\ }\href {https://doi.org/10.1103/PhysRevA.66.063609} {\bibfield
  {journal} {\bibinfo  {journal} {Phys. Rev. A}\ }\textbf {\bibinfo {volume}
  {66}},\ \bibinfo {pages} {063609} (\bibinfo {year} {2002})}\BibitemShut
  {NoStop}%
\bibitem [{\citenamefont {Cotner}(2016)}]{PhysRevD.94.063503}%
  \BibitemOpen
  \bibfield  {author} {\bibinfo {author} {\bibfnamefont {E.}~\bibnamefont
  {Cotner}},\ }\href {https://doi.org/10.1103/PhysRevD.94.063503} {\bibfield
  {journal} {\bibinfo  {journal} {Phys. Rev. D}\ }\textbf {\bibinfo {volume}
  {94}},\ \bibinfo {pages} {063503} (\bibinfo {year} {2016})}\BibitemShut
  {NoStop}%
\bibitem [{\citenamefont {{Gonz{\'a}lez}}\ and\ \citenamefont
  {{Guzm{\'a}n}}(2011)}]{2011PhRvD..83j3513G}%
  \BibitemOpen
  \bibfield  {author} {\bibinfo {author} {\bibfnamefont {J.~A.}\ \bibnamefont
  {{Gonz{\'a}lez}}}\ and\ \bibinfo {author} {\bibfnamefont {F.~S.}\
  \bibnamefont {{Guzm{\'a}n}}},\ }\href
  {https://doi.org/10.1103/PhysRevD.83.103513} {\bibfield  {journal} {\bibinfo
  {journal} {\prd}\ }\textbf {\bibinfo {volume} {83}},\ \bibinfo {eid} {103513}
  (\bibinfo {year} {2011})},\ \Eprint {https://arxiv.org/abs/1105.2066}
  {arXiv:1105.2066 [astro-ph.CO]} \BibitemShut {NoStop}%
\bibitem [{\citenamefont {{Bernal}}\ and\ \citenamefont
  {{Guzm{\'a}n}}(2006)}]{2006PhRvD..74j3002B}%
  \BibitemOpen
  \bibfield  {author} {\bibinfo {author} {\bibfnamefont {A.}~\bibnamefont
  {{Bernal}}}\ and\ \bibinfo {author} {\bibfnamefont {F.~S.}\ \bibnamefont
  {{Guzm{\'a}n}}},\ }\href {https://doi.org/10.1103/PhysRevD.74.103002}
  {\bibfield  {journal} {\bibinfo  {journal} {\prd}\ }\textbf {\bibinfo
  {volume} {74}},\ \bibinfo {eid} {103002} (\bibinfo {year} {2006})},\ \Eprint
  {https://arxiv.org/abs/astro-ph/0610682} {arXiv:astro-ph/0610682 [astro-ph]}
  \BibitemShut {NoStop}%
\bibitem [{\citenamefont {{Choi}}(2002)}]{2002PhRvA..66f3609C}%
  \BibitemOpen
  \bibfield  {author} {\bibinfo {author} {\bibfnamefont {D.-I.}\ \bibnamefont
  {{Choi}}},\ }\href {https://doi.org/10.1103/PhysRevA.66.063609} {\bibfield
  {journal} {\bibinfo  {journal} {\pra}\ }\textbf {\bibinfo {volume} {66}},\
  \bibinfo {eid} {063609} (\bibinfo {year} {2002})}\BibitemShut {NoStop}%
\bibitem [{\citenamefont {Carrasco}\ \emph
  {et~al.}(2010{\natexlab{a}})\citenamefont {Carrasco}, \citenamefont {Gomez},
  \citenamefont {Verdugo}, \citenamefont {Lee}, \citenamefont {Diaz},
  \citenamefont {Bergmann}, \citenamefont {Turner}, \citenamefont {Miller},\
  and\ \citenamefont {West}}]{2010ApJ...715L.160C}%
  \BibitemOpen
  \bibfield  {author} {\bibinfo {author} {\bibfnamefont {E.~R.}\ \bibnamefont
  {Carrasco}}, \bibinfo {author} {\bibfnamefont {P.~L.}\ \bibnamefont {Gomez}},
  \bibinfo {author} {\bibfnamefont {T.}~\bibnamefont {Verdugo}}, \bibinfo
  {author} {\bibfnamefont {H.}~\bibnamefont {Lee}}, \bibinfo {author}
  {\bibfnamefont {R.}~\bibnamefont {Diaz}}, \bibinfo {author} {\bibfnamefont
  {M.}~\bibnamefont {Bergmann}}, \bibinfo {author} {\bibfnamefont {J.~E.~H.}\
  \bibnamefont {Turner}}, \bibinfo {author} {\bibfnamefont {B.~W.}\
  \bibnamefont {Miller}},\ and\ \bibinfo {author} {\bibfnamefont {M.~J.}\
  \bibnamefont {West}},\ }\href {https://doi.org/10.1088/2041-8205/715/2/L160}
  {\bibfield  {journal} {\bibinfo  {journal} {The Astrophysical Journal
  Letters}\ }\textbf {\bibinfo {volume} {715}},\ \bibinfo {pages} {L160}
  (\bibinfo {year} {2010}{\natexlab{a}})}\BibitemShut {NoStop}%
\bibitem [{\citenamefont {Harvey}\ \emph {et~al.}(2015)\citenamefont {Harvey},
  \citenamefont {Massey}, \citenamefont {Kitching}, \citenamefont {Taylor},\
  and\ \citenamefont {Tittley}}]{Harvey_2015}%
  \BibitemOpen
  \bibfield  {author} {\bibinfo {author} {\bibfnamefont {D.}~\bibnamefont
  {Harvey}}, \bibinfo {author} {\bibfnamefont {R.}~\bibnamefont {Massey}},
  \bibinfo {author} {\bibfnamefont {T.}~\bibnamefont {Kitching}}, \bibinfo
  {author} {\bibfnamefont {A.}~\bibnamefont {Taylor}},\ and\ \bibinfo {author}
  {\bibfnamefont {E.}~\bibnamefont {Tittley}},\ }\href
  {https://doi.org/10.1126/science.1261381} {\bibfield  {journal} {\bibinfo
  {journal} {Science}\ }\textbf {\bibinfo {volume} {347}},\ \bibinfo {pages}
  {1462} (\bibinfo {year} {2015})}\BibitemShut {NoStop}%
\bibitem [{\citenamefont {Paredes}\ and\ \citenamefont
  {Michinel}(2016{\natexlab{a}})}]{2016PDU....12...50P}%
  \BibitemOpen
  \bibfield  {author} {\bibinfo {author} {\bibfnamefont {A.}~\bibnamefont
  {Paredes}}\ and\ \bibinfo {author} {\bibfnamefont {H.}~\bibnamefont
  {Michinel}},\ }\href {https://doi.org/10.1016/j.dark.2016.02.003} {\bibfield
  {journal} {\bibinfo  {journal} {Physics of the Dark Universe}\ }\textbf
  {\bibinfo {volume} {12}},\ \bibinfo {pages} {50} (\bibinfo {year}
  {2016}{\natexlab{a}})},\ \Eprint {https://arxiv.org/abs/1512.05121}
  {arXiv:1512.05121 [astro-ph.CO]} \BibitemShut {NoStop}%
\bibitem [{\citenamefont {{Guzm{\'a}n}}\ and\ \citenamefont
  {{Avilez}}(2018)}]{2018PhRvD..97k6003G}%
  \BibitemOpen
  \bibfield  {author} {\bibinfo {author} {\bibfnamefont {F.~S.}\ \bibnamefont
  {{Guzm{\'a}n}}}\ and\ \bibinfo {author} {\bibfnamefont {A.~A.}\ \bibnamefont
  {{Avilez}}},\ }\href {https://doi.org/10.1103/PhysRevD.97.116003} {\bibfield
  {journal} {\bibinfo  {journal} {\prd}\ }\textbf {\bibinfo {volume} {97}},\
  \bibinfo {eid} {116003} (\bibinfo {year} {2018})},\ \Eprint
  {https://arxiv.org/abs/1804.08670} {arXiv:1804.08670 [gr-qc]} \BibitemShut
  {NoStop}%
\bibitem [{\citenamefont {{Schwabe}}\ \emph {et~al.}(2016)\citenamefont
  {{Schwabe}}, \citenamefont {{Niemeyer}},\ and\ \citenamefont
  {{Engels}}}]{2016PhRvD..94d3513S}%
  \BibitemOpen
  \bibfield  {author} {\bibinfo {author} {\bibfnamefont {B.}~\bibnamefont
  {{Schwabe}}}, \bibinfo {author} {\bibfnamefont {J.~C.}\ \bibnamefont
  {{Niemeyer}}},\ and\ \bibinfo {author} {\bibfnamefont {J.~F.}\ \bibnamefont
  {{Engels}}},\ }\href {https://doi.org/10.1103/PhysRevD.94.043513} {\bibfield
  {journal} {\bibinfo  {journal} {\prd}\ }\textbf {\bibinfo {volume} {94}},\
  \bibinfo {eid} {043513} (\bibinfo {year} {2016})},\ \Eprint
  {https://arxiv.org/abs/1606.05151} {arXiv:1606.05151 [astro-ph.CO]}
  \BibitemShut {NoStop}%
\bibitem [{\citenamefont {Liu}\ \emph {et~al.}(2022)\citenamefont {Liu},
  \citenamefont {Proukakis}, \citenamefont {Rigopoulos} \emph
  {et~al.}}]{liu2022coherent}%
  \BibitemOpen
  \bibfield  {author} {\bibinfo {author} {\bibfnamefont {I.}~\bibnamefont
  {Liu}}, \bibinfo {author} {\bibfnamefont {N.~P.}\ \bibnamefont {Proukakis}},
  \bibinfo {author} {\bibfnamefont {G.}~\bibnamefont {Rigopoulos}}, \emph
  {et~al.},\ }\href@noop {} {\bibfield  {journal} {\bibinfo  {journal} {arXiv
  preprint arXiv:2211.02565}\ } (\bibinfo {year} {2022})}\BibitemShut {NoStop}%
\bibitem [{\citenamefont {{Schive}}\ \emph
  {et~al.}(2014{\natexlab{b}})\citenamefont {{Schive}}, \citenamefont {{Liao}},
  \citenamefont {{Woo}}, \citenamefont {{Wong}}, \citenamefont {{Chiueh}},
  \citenamefont {{Broadhurst}},\ and\ \citenamefont
  {{Hwang}}}]{2014PhRvL.113z1302S}%
  \BibitemOpen
  \bibfield  {author} {\bibinfo {author} {\bibfnamefont {H.-Y.}\ \bibnamefont
  {{Schive}}}, \bibinfo {author} {\bibfnamefont {M.-H.}\ \bibnamefont
  {{Liao}}}, \bibinfo {author} {\bibfnamefont {T.-P.}\ \bibnamefont {{Woo}}},
  \bibinfo {author} {\bibfnamefont {S.-K.}\ \bibnamefont {{Wong}}}, \bibinfo
  {author} {\bibfnamefont {T.}~\bibnamefont {{Chiueh}}}, \bibinfo {author}
  {\bibfnamefont {T.}~\bibnamefont {{Broadhurst}}},\ and\ \bibinfo {author}
  {\bibfnamefont {W.~Y.~P.}\ \bibnamefont {{Hwang}}},\ }\href
  {https://doi.org/10.1103/PhysRevLett.113.261302} {\bibfield  {journal}
  {\bibinfo  {journal} {\prl}\ }\textbf {\bibinfo {volume} {113}},\ \bibinfo
  {eid} {261302} (\bibinfo {year} {2014}{\natexlab{b}})},\ \Eprint
  {https://arxiv.org/abs/1407.7762} {arXiv:1407.7762 [astro-ph.GA]}
  \BibitemShut {NoStop}%
\bibitem [{\citenamefont {Maleki}\ \emph {et~al.}(2020)\citenamefont {Maleki},
  \citenamefont {Baghram},\ and\ \citenamefont {Rahvar}}]{Maleki_2020}%
  \BibitemOpen
  \bibfield  {author} {\bibinfo {author} {\bibfnamefont {A.}~\bibnamefont
  {Maleki}}, \bibinfo {author} {\bibfnamefont {S.}~\bibnamefont {Baghram}},\
  and\ \bibinfo {author} {\bibfnamefont {S.}~\bibnamefont {Rahvar}},\
  }\bibfield  {journal} {\bibinfo  {journal} {Physical Review D}\ }\textbf
  {\bibinfo {volume} {101}},\ \href
  {https://doi.org/10.1103/physrevd.101.023508} {10.1103/physrevd.101.023508}
  (\bibinfo {year} {2020})\BibitemShut {NoStop}%
\bibitem [{\citenamefont {Paredes}\ and\ \citenamefont
  {Michinel}(2016{\natexlab{b}})}]{Paredes_2016}%
  \BibitemOpen
  \bibfield  {author} {\bibinfo {author} {\bibfnamefont {A.}~\bibnamefont
  {Paredes}}\ and\ \bibinfo {author} {\bibfnamefont {H.}~\bibnamefont
  {Michinel}},\ }\href {https://doi.org/10.1016/j.dark.2016.02.003} {\bibfield
  {journal} {\bibinfo  {journal} {Physics of the Dark Universe}\ }\textbf
  {\bibinfo {volume} {12}},\ \bibinfo {pages} {50} (\bibinfo {year}
  {2016}{\natexlab{b}})}\BibitemShut {NoStop}%
\bibitem [{\citenamefont {Carrasco}\ \emph
  {et~al.}(2010{\natexlab{b}})\citenamefont {Carrasco}, \citenamefont {Gomez},
  \citenamefont {Verdugo}, \citenamefont {Lee}, \citenamefont {Diaz},
  \citenamefont {Bergmann}, \citenamefont {Turner}, \citenamefont {Miller},\
  and\ \citenamefont {West}}]{Carrasco_2010}%
  \BibitemOpen
  \bibfield  {author} {\bibinfo {author} {\bibfnamefont {E.~R.}\ \bibnamefont
  {Carrasco}}, \bibinfo {author} {\bibfnamefont {P.~L.}\ \bibnamefont {Gomez}},
  \bibinfo {author} {\bibfnamefont {T.}~\bibnamefont {Verdugo}}, \bibinfo
  {author} {\bibfnamefont {H.}~\bibnamefont {Lee}}, \bibinfo {author}
  {\bibfnamefont {R.}~\bibnamefont {Diaz}}, \bibinfo {author} {\bibfnamefont
  {M.}~\bibnamefont {Bergmann}}, \bibinfo {author} {\bibfnamefont {J.~E.~H.}\
  \bibnamefont {Turner}}, \bibinfo {author} {\bibfnamefont {B.~W.}\
  \bibnamefont {Miller}},\ and\ \bibinfo {author} {\bibfnamefont {M.~J.}\
  \bibnamefont {West}},\ }\href {https://doi.org/10.1088/2041-8205/715/2/L160}
  {\bibfield  {journal} {\bibinfo  {journal} {The Astrophysical Journal
  Letters}\ }\textbf {\bibinfo {volume} {715}},\ \bibinfo {pages} {L160}
  (\bibinfo {year} {2010}{\natexlab{b}})}\BibitemShut {NoStop}%
\bibitem [{\citenamefont {Massey}\ \emph {et~al.}(2015)\citenamefont {Massey},
  \citenamefont {Williams}, \citenamefont {Smit}, \citenamefont {Swinbank},
  \citenamefont {Kitching}, \citenamefont {Harvey}, \citenamefont {Jauzac},
  \citenamefont {Israel}, \citenamefont {Clowe}, \citenamefont {Edge},
  \citenamefont {Hilton}, \citenamefont {Jullo}, \citenamefont {Leonard},
  \citenamefont {Liesenborgs}, \citenamefont {Merten}, \citenamefont
  {Mohammed}, \citenamefont {Nagai}, \citenamefont {Richard}, \citenamefont
  {Robertson}, \citenamefont {Saha}, \citenamefont {Santana}, \citenamefont
  {Stott},\ and\ \citenamefont {Tittley}}]{Massey_2015}%
  \BibitemOpen
  \bibfield  {author} {\bibinfo {author} {\bibfnamefont {R.}~\bibnamefont
  {Massey}}, \bibinfo {author} {\bibfnamefont {L.}~\bibnamefont {Williams}},
  \bibinfo {author} {\bibfnamefont {R.}~\bibnamefont {Smit}}, \bibinfo {author}
  {\bibfnamefont {M.}~\bibnamefont {Swinbank}}, \bibinfo {author}
  {\bibfnamefont {T.~D.}\ \bibnamefont {Kitching}}, \bibinfo {author}
  {\bibfnamefont {D.}~\bibnamefont {Harvey}}, \bibinfo {author} {\bibfnamefont
  {M.}~\bibnamefont {Jauzac}}, \bibinfo {author} {\bibfnamefont
  {H.}~\bibnamefont {Israel}}, \bibinfo {author} {\bibfnamefont
  {D.}~\bibnamefont {Clowe}}, \bibinfo {author} {\bibfnamefont
  {A.}~\bibnamefont {Edge}}, \bibinfo {author} {\bibfnamefont {M.}~\bibnamefont
  {Hilton}}, \bibinfo {author} {\bibfnamefont {E.}~\bibnamefont {Jullo}},
  \bibinfo {author} {\bibfnamefont {A.}~\bibnamefont {Leonard}}, \bibinfo
  {author} {\bibfnamefont {J.}~\bibnamefont {Liesenborgs}}, \bibinfo {author}
  {\bibfnamefont {J.}~\bibnamefont {Merten}}, \bibinfo {author} {\bibfnamefont
  {I.}~\bibnamefont {Mohammed}}, \bibinfo {author} {\bibfnamefont
  {D.}~\bibnamefont {Nagai}}, \bibinfo {author} {\bibfnamefont
  {J.}~\bibnamefont {Richard}}, \bibinfo {author} {\bibfnamefont
  {A.}~\bibnamefont {Robertson}}, \bibinfo {author} {\bibfnamefont
  {P.}~\bibnamefont {Saha}}, \bibinfo {author} {\bibfnamefont {R.}~\bibnamefont
  {Santana}}, \bibinfo {author} {\bibfnamefont {J.}~\bibnamefont {Stott}},\
  and\ \bibinfo {author} {\bibfnamefont {E.}~\bibnamefont {Tittley}},\ }\href
  {https://doi.org/10.1093/mnras/stv467} {\bibfield  {journal} {\bibinfo
  {journal} {Monthly Notices of the Royal Astronomical Society}\ }\textbf
  {\bibinfo {volume} {449}},\ \bibinfo {pages} {3393} (\bibinfo {year}
  {2015})}\BibitemShut {NoStop}%
\bibitem [{\citenamefont {{Avilez}}\ and\ \citenamefont
  {{Guzm{\'a}n}}(2019)}]{2019PhRvD..99d3542A}%
  \BibitemOpen
  \bibfield  {author} {\bibinfo {author} {\bibfnamefont {A.~A.}\ \bibnamefont
  {{Avilez}}}\ and\ \bibinfo {author} {\bibfnamefont {F.~S.}\ \bibnamefont
  {{Guzm{\'a}n}}},\ }\href {https://doi.org/10.1103/PhysRevD.99.043542}
  {\bibfield  {journal} {\bibinfo  {journal} {\prd}\ }\textbf {\bibinfo
  {volume} {99}},\ \bibinfo {eid} {043542} (\bibinfo {year}
  {2019})}\BibitemShut {NoStop}%
\bibitem [{\citenamefont {Guzm\'an}\ \emph {et~al.}(2016)\citenamefont
  {Guzm\'an}, \citenamefont {Gonz\'alez},\ and\ \citenamefont
  {Cruz-P\'erez}}]{PhysRevD.93.103535}%
  \BibitemOpen
  \bibfield  {author} {\bibinfo {author} {\bibfnamefont {F.~S.}\ \bibnamefont
  {Guzm\'an}}, \bibinfo {author} {\bibfnamefont {J.~A.}\ \bibnamefont
  {Gonz\'alez}},\ and\ \bibinfo {author} {\bibfnamefont {J.~P.}\ \bibnamefont
  {Cruz-P\'erez}},\ }\href {https://doi.org/10.1103/PhysRevD.93.103535}
  {\bibfield  {journal} {\bibinfo  {journal} {Phys. Rev. D}\ }\textbf {\bibinfo
  {volume} {93}},\ \bibinfo {pages} {103535} (\bibinfo {year}
  {2016})}\BibitemShut {NoStop}%
\bibitem [{\citenamefont {{Palenzuela}}\ \emph {et~al.}(2007)\citenamefont
  {{Palenzuela}}, \citenamefont {{Olabarrieta}}, \citenamefont {{Lehner}},\
  and\ \citenamefont {{Liebling}}}]{2007PhRvD..75f4005P}%
  \BibitemOpen
  \bibfield  {author} {\bibinfo {author} {\bibfnamefont {C.}~\bibnamefont
  {{Palenzuela}}}, \bibinfo {author} {\bibfnamefont {I.}~\bibnamefont
  {{Olabarrieta}}}, \bibinfo {author} {\bibfnamefont {L.}~\bibnamefont
  {{Lehner}}},\ and\ \bibinfo {author} {\bibfnamefont {S.~L.}\ \bibnamefont
  {{Liebling}}},\ }\href {https://doi.org/10.1103/PhysRevD.75.064005}
  {\bibfield  {journal} {\bibinfo  {journal} {\prd}\ }\textbf {\bibinfo
  {volume} {75}},\ \bibinfo {eid} {064005} (\bibinfo {year} {2007})},\ \Eprint
  {https://arxiv.org/abs/gr-qc/0612067} {arXiv:gr-qc/0612067 [gr-qc]}
  \BibitemShut {NoStop}%
\bibitem [{\citenamefont {{Palenzuela}}\ \emph {et~al.}(2008)\citenamefont
  {{Palenzuela}}, \citenamefont {{Lehner}},\ and\ \citenamefont
  {{Liebling}}}]{2008PhRvD..77d4036P}%
  \BibitemOpen
  \bibfield  {author} {\bibinfo {author} {\bibfnamefont {C.}~\bibnamefont
  {{Palenzuela}}}, \bibinfo {author} {\bibfnamefont {L.}~\bibnamefont
  {{Lehner}}},\ and\ \bibinfo {author} {\bibfnamefont {S.~L.}\ \bibnamefont
  {{Liebling}}},\ }\href {https://doi.org/10.1103/PhysRevD.77.044036}
  {\bibfield  {journal} {\bibinfo  {journal} {\prd}\ }\textbf {\bibinfo
  {volume} {77}},\ \bibinfo {eid} {044036} (\bibinfo {year} {2008})},\ \Eprint
  {https://arxiv.org/abs/0706.2435} {arXiv:0706.2435 [gr-qc]} \BibitemShut
  {NoStop}%
\bibitem [{\citenamefont {{Bezares}}\ \emph {et~al.}(2017)\citenamefont
  {{Bezares}}, \citenamefont {{Palenzuela}},\ and\ \citenamefont
  {{Bona}}}]{2017PhRvD..95l4005B}%
  \BibitemOpen
  \bibfield  {author} {\bibinfo {author} {\bibfnamefont {M.}~\bibnamefont
  {{Bezares}}}, \bibinfo {author} {\bibfnamefont {C.}~\bibnamefont
  {{Palenzuela}}},\ and\ \bibinfo {author} {\bibfnamefont {C.}~\bibnamefont
  {{Bona}}},\ }\href {https://doi.org/10.1103/PhysRevD.95.124005} {\bibfield
  {journal} {\bibinfo  {journal} {\prd}\ }\textbf {\bibinfo {volume} {95}},\
  \bibinfo {eid} {124005} (\bibinfo {year} {2017})},\ \Eprint
  {https://arxiv.org/abs/1705.01071} {arXiv:1705.01071 [gr-qc]} \BibitemShut
  {NoStop}%
\bibitem [{\citenamefont {{Balakrishna}}(1999)}]{1999PhDT........44B}%
  \BibitemOpen
  \bibfield  {author} {\bibinfo {author} {\bibfnamefont {J.}~\bibnamefont
  {{Balakrishna}}},\ }\emph {\bibinfo {title} {{A numerical study of boson
  stars: Einstein equations with a matter source}}},\ \href@noop {} {Ph.D.
  thesis},\ \bibinfo  {school} {Washington University in Saint Louis, Missouri}
  (\bibinfo {year} {1999})\BibitemShut {NoStop}%
\bibitem [{\citenamefont {{Choptuik}}\ and\ \citenamefont
  {{Pretorius}}(2010)}]{2010PhRvL.104k1101C}%
  \BibitemOpen
  \bibfield  {author} {\bibinfo {author} {\bibfnamefont {M.~W.}\ \bibnamefont
  {{Choptuik}}}\ and\ \bibinfo {author} {\bibfnamefont {F.}~\bibnamefont
  {{Pretorius}}},\ }\href {https://doi.org/10.1103/PhysRevLett.104.111101}
  {\bibfield  {journal} {\bibinfo  {journal} {\prl}\ }\textbf {\bibinfo
  {volume} {104}},\ \bibinfo {eid} {111101} (\bibinfo {year} {2010})},\ \Eprint
  {https://arxiv.org/abs/0908.1780} {arXiv:0908.1780 [gr-qc]} \BibitemShut
  {NoStop}%
\bibitem [{\citenamefont {{Guzm{\'a}n}}\ \emph {et~al.}(2016)\citenamefont
  {{Guzm{\'a}n}}, \citenamefont {{Gonz{\'a}lez}},\ and\ \citenamefont
  {{Cruz-P{\'e}rez}}}]{2016PhRvD..93j3535G}%
  \BibitemOpen
  \bibfield  {author} {\bibinfo {author} {\bibfnamefont {F.~S.}\ \bibnamefont
  {{Guzm{\'a}n}}}, \bibinfo {author} {\bibfnamefont {J.~A.}\ \bibnamefont
  {{Gonz{\'a}lez}}},\ and\ \bibinfo {author} {\bibfnamefont {J.~P.}\
  \bibnamefont {{Cruz-P{\'e}rez}}},\ }\href
  {https://doi.org/10.1103/PhysRevD.93.103535} {\bibfield  {journal} {\bibinfo
  {journal} {\prd}\ }\textbf {\bibinfo {volume} {93}},\ \bibinfo {eid} {103535}
  (\bibinfo {year} {2016})},\ \Eprint {https://arxiv.org/abs/1605.04856}
  {arXiv:1605.04856 [astro-ph.GA]} \BibitemShut {NoStop}%
\bibitem [{\citenamefont {{Markevitch}}\ \emph {et~al.}(2004)\citenamefont
  {{Markevitch}}, \citenamefont {{Gonzalez}}, \citenamefont {{Clowe}},
  \citenamefont {{Vikhlinin}}, \citenamefont {{Forman}}, \citenamefont
  {{Jones}}, \citenamefont {{Murray}},\ and\ \citenamefont
  {{Tucker}}}]{2004ApJ...606..819M}%
  \BibitemOpen
  \bibfield  {author} {\bibinfo {author} {\bibfnamefont {M.}~\bibnamefont
  {{Markevitch}}}, \bibinfo {author} {\bibfnamefont {A.~H.}\ \bibnamefont
  {{Gonzalez}}}, \bibinfo {author} {\bibfnamefont {D.}~\bibnamefont {{Clowe}}},
  \bibinfo {author} {\bibfnamefont {A.}~\bibnamefont {{Vikhlinin}}}, \bibinfo
  {author} {\bibfnamefont {W.}~\bibnamefont {{Forman}}}, \bibinfo {author}
  {\bibfnamefont {C.}~\bibnamefont {{Jones}}}, \bibinfo {author} {\bibfnamefont
  {S.}~\bibnamefont {{Murray}}},\ and\ \bibinfo {author} {\bibfnamefont
  {W.}~\bibnamefont {{Tucker}}},\ }\href {https://doi.org/10.1086/383178}
  {\bibfield  {journal} {\bibinfo  {journal} {\apj}\ }\textbf {\bibinfo
  {volume} {606}},\ \bibinfo {pages} {819} (\bibinfo {year} {2004})},\ \Eprint
  {https://arxiv.org/abs/astro-ph/0309303} {arXiv:astro-ph/0309303 [astro-ph]}
  \BibitemShut {NoStop}%
\bibitem [{\citenamefont {{Liebling}}\ and\ \citenamefont
  {{Palenzuela}}(2012)}]{2012LRR....15....6L}%
  \BibitemOpen
  \bibfield  {author} {\bibinfo {author} {\bibfnamefont {S.~L.}\ \bibnamefont
  {{Liebling}}}\ and\ \bibinfo {author} {\bibfnamefont {C.}~\bibnamefont
  {{Palenzuela}}},\ }\href {https://doi.org/10.12942/lrr-2012-6} {\bibfield
  {journal} {\bibinfo  {journal} {Living Reviews in Relativity}\ }\textbf
  {\bibinfo {volume} {15}},\ \bibinfo {eid} {6} (\bibinfo {year} {2012})},\
  \Eprint {https://arxiv.org/abs/1202.5809} {arXiv:1202.5809 [gr-qc]}
  \BibitemShut {NoStop}%
\bibitem [{\citenamefont {{Macedo}}\ \emph {et~al.}(2013)\citenamefont
  {{Macedo}}, \citenamefont {{Pani}}, \citenamefont {{Cardoso}},\ and\
  \citenamefont {{Crispino}}}]{2013PhRvD..88f4046M}%
  \BibitemOpen
  \bibfield  {author} {\bibinfo {author} {\bibfnamefont {C.~F.~B.}\
  \bibnamefont {{Macedo}}}, \bibinfo {author} {\bibfnamefont {P.}~\bibnamefont
  {{Pani}}}, \bibinfo {author} {\bibfnamefont {V.}~\bibnamefont {{Cardoso}}},\
  and\ \bibinfo {author} {\bibfnamefont {L.~C.~B.}\ \bibnamefont
  {{Crispino}}},\ }\href {https://doi.org/10.1103/PhysRevD.88.064046}
  {\bibfield  {journal} {\bibinfo  {journal} {\prd}\ }\textbf {\bibinfo
  {volume} {88}},\ \bibinfo {eid} {064046} (\bibinfo {year} {2013})},\ \Eprint
  {https://arxiv.org/abs/1307.4812} {arXiv:1307.4812 [gr-qc]} \BibitemShut
  {NoStop}%
\bibitem [{\citenamefont {Antoine}\ \emph {et~al.}(2013)\citenamefont
  {Antoine}, \citenamefont {Bao},\ and\ \citenamefont
  {Besse}}]{ANTOINE20132621}%
  \BibitemOpen
  \bibfield  {author} {\bibinfo {author} {\bibfnamefont {X.}~\bibnamefont
  {Antoine}}, \bibinfo {author} {\bibfnamefont {W.}~\bibnamefont {Bao}},\ and\
  \bibinfo {author} {\bibfnamefont {C.}~\bibnamefont {Besse}},\ }\href
  {https://doi.org/https://doi.org/10.1016/j.cpc.2013.07.012} {\bibfield
  {journal} {\bibinfo  {journal} {Computer Physics Communications}\ }\textbf
  {\bibinfo {volume} {184}},\ \bibinfo {pages} {2621} (\bibinfo {year}
  {2013})}\BibitemShut {NoStop}%
\bibitem [{\citenamefont {Costiner}\ and\ \citenamefont
  {Ta'asan}(1995)}]{PhysRevE.52.1181}%
  \BibitemOpen
  \bibfield  {author} {\bibinfo {author} {\bibfnamefont {S.}~\bibnamefont
  {Costiner}}\ and\ \bibinfo {author} {\bibfnamefont {S.}~\bibnamefont
  {Ta'asan}},\ }\href {https://doi.org/10.1103/PhysRevE.52.1181} {\bibfield
  {journal} {\bibinfo  {journal} {Phys. Rev. E}\ }\textbf {\bibinfo {volume}
  {52}},\ \bibinfo {pages} {1181} (\bibinfo {year} {1995})}\BibitemShut
  {NoStop}%
\bibitem [{\citenamefont {Munive-Villa}\ \emph {et~al.}(2022)\citenamefont
  {Munive-Villa}, \citenamefont {L\'opez-S\'anchez}, \citenamefont
  {Avilez-L\'opez},\ and\ \citenamefont {Guzm\'an}}]{PhysRevD.105.083521}%
  \BibitemOpen
  \bibfield  {author} {\bibinfo {author} {\bibfnamefont {E.}~\bibnamefont
  {Munive-Villa}}, \bibinfo {author} {\bibfnamefont {J.~N.}\ \bibnamefont
  {L\'opez-S\'anchez}}, \bibinfo {author} {\bibfnamefont {A.~A.}\ \bibnamefont
  {Avilez-L\'opez}},\ and\ \bibinfo {author} {\bibfnamefont {F.~S.}\
  \bibnamefont {Guzm\'an}},\ }\href
  {https://doi.org/10.1103/PhysRevD.105.083521} {\bibfield  {journal} {\bibinfo
   {journal} {Phys. Rev. D}\ }\textbf {\bibinfo {volume} {105}},\ \bibinfo
  {pages} {083521} (\bibinfo {year} {2022})}\BibitemShut {NoStop}%
\bibitem [{\citenamefont {O’Reilly}\ and\ \citenamefont
  {Beck}(2006)}]{OReilly2006AFO}%
  \BibitemOpen
  \bibfield  {author} {\bibinfo {author} {\bibfnamefont {R.~C.}\ \bibnamefont
  {O’Reilly}}\ and\ \bibinfo {author} {\bibfnamefont {J.~M.}\ \bibnamefont
  {Beck}}\ }(\bibinfo {year} {2006})\BibitemShut {NoStop}%
\bibitem [{\citenamefont {{Bazhan}}\ \emph {et~al.}(2022)\citenamefont
  {{Bazhan}}, \citenamefont {{Svetlichnyi}}, \citenamefont {{Pfeiffer}},
  \citenamefont {{Derr}}, \citenamefont {{Birkl}},\ and\ \citenamefont
  {{Yakimenko}}}]{2022PhRvA.106d3305B}%
  \BibitemOpen
  \bibfield  {author} {\bibinfo {author} {\bibfnamefont {N.}~\bibnamefont
  {{Bazhan}}}, \bibinfo {author} {\bibfnamefont {A.}~\bibnamefont
  {{Svetlichnyi}}}, \bibinfo {author} {\bibfnamefont {D.}~\bibnamefont
  {{Pfeiffer}}}, \bibinfo {author} {\bibfnamefont {D.}~\bibnamefont {{Derr}}},
  \bibinfo {author} {\bibfnamefont {G.}~\bibnamefont {{Birkl}}},\ and\ \bibinfo
  {author} {\bibfnamefont {A.}~\bibnamefont {{Yakimenko}}},\ }\href
  {https://doi.org/10.1103/PhysRevA.106.043305} {\bibfield  {journal} {\bibinfo
   {journal} {\pra}\ }\textbf {\bibinfo {volume} {106}},\ \bibinfo {eid}
  {043305} (\bibinfo {year} {2022})},\ \Eprint
  {https://arxiv.org/abs/2204.14269} {arXiv:2204.14269 [cond-mat.quant-gas]}
  \BibitemShut {NoStop}%
\bibitem [{\citenamefont {Poisson}\ and\ \citenamefont
  {Will}(2014)}]{poisson2014gravity}%
  \BibitemOpen
  \bibfield  {author} {\bibinfo {author} {\bibfnamefont {E.}~\bibnamefont
  {Poisson}}\ and\ \bibinfo {author} {\bibfnamefont {C.~M.}\ \bibnamefont
  {Will}},\ }\href@noop {} {\emph {\bibinfo {title} {Gravity: Newtonian,
  post-newtonian, relativistic}}}\ (\bibinfo  {publisher} {Cambridge University
  Press},\ \bibinfo {year} {2014})\BibitemShut {NoStop}%
\bibitem [{\citenamefont {Quilis}\ \emph {et~al.}(2007)\citenamefont {Quilis},
  \citenamefont {Gonz\'alez-Garc\'{\i}a}, \citenamefont {S\'aez},\ and\
  \citenamefont {Font}}]{quilis}%
  \BibitemOpen
  \bibfield  {author} {\bibinfo {author} {\bibfnamefont {V.}~\bibnamefont
  {Quilis}}, \bibinfo {author} {\bibfnamefont {A.~C.}\ \bibnamefont
  {Gonz\'alez-Garc\'{\i}a}}, \bibinfo {author} {\bibfnamefont {D.}~\bibnamefont
  {S\'aez}},\ and\ \bibinfo {author} {\bibfnamefont {J.~A.}\ \bibnamefont
  {Font}},\ }\href {https://doi.org/10.1103/PhysRevD.75.104008} {\bibfield
  {journal} {\bibinfo  {journal} {Phys. Rev. D}\ }\textbf {\bibinfo {volume}
  {75}},\ \bibinfo {pages} {104008} (\bibinfo {year} {2007})}\BibitemShut
  {NoStop}%
\bibitem [{\citenamefont {Inagaki}\ \emph {et~al.}(2010)\citenamefont
  {Inagaki}, \citenamefont {Takahashi}, \citenamefont {Masaki},\ and\
  \citenamefont {Sugiyama}}]{Inagaki:2010nu}%
  \BibitemOpen
  \bibfield  {author} {\bibinfo {author} {\bibfnamefont {T.}~\bibnamefont
  {Inagaki}}, \bibinfo {author} {\bibfnamefont {K.}~\bibnamefont {Takahashi}},
  \bibinfo {author} {\bibfnamefont {S.}~\bibnamefont {Masaki}},\ and\ \bibinfo
  {author} {\bibfnamefont {N.}~\bibnamefont {Sugiyama}},\ }\href
  {https://doi.org/10.1103/PhysRevD.82.124007} {\bibfield  {journal} {\bibinfo
  {journal} {Phys. Rev. D}\ }\textbf {\bibinfo {volume} {82}},\ \bibinfo
  {pages} {124007} (\bibinfo {year} {2010})},\ \Eprint
  {https://arxiv.org/abs/1011.5554} {arXiv:1011.5554 [astro-ph.CO]}
  \BibitemShut {NoStop}%
\end{thebibliography}%

\end{document}